\documentclass[a4paper,amsmath,amssymb,twocolumn,11pt]{quantumarticle}
\pdfoutput=1
\usepackage[utf8]{inputenc}
\usepackage[english]{babel}
\usepackage[T1]{fontenc}
\usepackage{hyperref}
\usepackage{bm}
\usepackage[numbers]{natbib}
\usepackage{braket}
\usepackage{enumitem}
\usepackage[normalem]{ulem}
\usepackage[caption=false]{subfig}

\newcommand{\widebar}[1]{\overline{#1}}

\newcommand{\rvec}[2]{\begin{pmatrix} #1 & #2 \end{pmatrix}} 
\newcommand{\mat}[2]{\begin{pmatrix} #1 \\ #2 \end{pmatrix}} 
\newcommand{\deriv}[2]{\frac{{\rm d}#1}{{\rm d}#2}}

\newcommand{\dmu}{\delta\mu}
\newcommand{\nuz}{\nu z}
\newcommand{\mcE}{\mathcal{E}}
\newcommand{\bmcE}{\widebar{\mathcal{E}}} 

\newcommand{\pdag}{{\phantom{\dagger}}}
\newcommand{\ave}[1]{\langle #1 \rangle}

\newcommand{\zetaA}{{\zeta_{A}}}
\newcommand{\zetaB}{{\zeta_{B}}}
\newcommand{\zetaC}{{\zeta_{C}}}
\newcommand{\zetaCoh}{{\zeta_{\rm coh}}}

\renewcommand{\theenumi}{(\roman{enumi})}

\begin{document}

\title{Nonlinear quantum Kibble-Zurek ramps in open systems at finite temperature}

\author{Johannes N.~Kriel}
\email{hkriel@sun.ac.za} 
\affiliation{Institute of Theoretical Physics, Stellenbosch University, Stellenbosch 7600, South Africa}

\author{Emma C.~King}
\affiliation{Theoretische Physik, Universität des Saarlandes, 66123 Saarbrücken, Germany}

\author{Michael Kastner}
\affiliation{Hanse-Wissenschaftskolleg, Lehmkuhlenbusch 4, 27753 Delmenhorst, Germany}

\maketitle

\begin{abstract}
We analyze quantum systems under a broad class of protocols in which the temperature and a Hamiltonian control parameter are ramped simultaneously and, in general, in a nonlinear fashion toward a quantum critical point. Using an open-system version of a Kitaev quantum wire as an example, we show that, unlike finite-temperature protocols at fixed temperature, these protocols allow us to probe, in an out-of-equilibrium situation and at finite temperature, the universality class (characterized by the critical exponents $\nu$ and $z$) of an equilibrium quantum phase transition at zero temperature. Key to this is the identification of ramps in which both coherent and incoherent parts of the open-system dynamics affect the excitation density in a non-negligible way. We also identify the specific ramps for which subleading corrections to the asymptotic scaling laws are suppressed, which serves as a guide to dynamically probing quantum critical exponents in experimentally realistic finite-temperature situations.
\end{abstract}

Scaling and universality have been known to occur in the vicinity of continuous equilibrium phase transitions for a long time, with the first experimental observations dating back to at least the 1940s \cite{Guggenheim45}. A theoretical framework for understanding and deriving scaling and universality from the underlying microscopic theory was developed in the subsequent decades, culminating in the renormalization group approach to equilibrium critical phenomena \cite{Kadanoff66,Wilson75,Cardy}. Out of equilibrium, occurrences of scaling and universality have been reported as well, but, while several promising ideas have been put forward \cite{Hinrichsen00,Odor04,BergesRothkopfSchmidt08,Heyl15}, much less is known and a general framework is lacking.

A popular and much-studied strategy for generating scaling behavior and universality in nonequilibrium situations is by devising protocols that drive a system out of equilibrium in such a way that the critical behavior of an underlying {\em equilibrium} phase transition leaves an imprint in nonequilibrium quantities. A prime example of this strategy goes by the name of ``quantum Kibble-Zurek mechanism'' \cite{ZurekDornerZoller05,Polkovnikov05,schutzhold2006}: For a system in equilibrium at zero temperature and at some initial parameter value $\mu$ sufficiently far from its quantum critical value $\mu_c$, a gradual change of $\mu$ results in adiabatic dynamics and leaves the system equilibrated. Only when $\mu$ gets sufficiently close to $\mu_c$, critical slowing down, i.e., the power-law divergence of the relaxation time of a many-body system in the vicinity of a continuous phase transition, prevents further adiabatic evolution and causes an approximate ``freeze-out'': The system's evolution becomes negligible in comparison to the timescale induced by the external driving of the system, which forces the system out of equilibrium. Suitably chosen nonequilibrium quantities then display universality and scaling laws governed by the critical exponents of the underlying equilibrium quantum phase transition. The rationale behind the Kibble-Zurek protocol, and also the ``art of the game'' more generally, is to devise nonequilibrium protocols and identify suitable observables that not only encode the equilibrium critical exponents, but ideally do so in the form of simple, clean scaling laws.

While, strictly speaking, quantum phase transitions occur at zero temperature, actual experiments are conducted at nonzero temperatures. However, it is well known that the presence of a quantum phase transition at zero temperatures leaves an imprint also in physical quantities at low,  nonzero temperatures. In the context of quantum Kibble-Zurek protocols and the probing of scaling and universality in nonequilibrium situations, it may therefore seem desirable to devise protocols that operate at nonzero temperatures, but nonetheless lead to scaling laws governed by the universal critical exponents or scaling functions of the underlying quantum critical point. With this goal in mind, a generalization of the above-described quantum Kibble-Zurek protocol has been analyzed by Patan\`e {\em et al.}\ in Refs.~\cite{Patane_etal08,Patane_etal09} and Yin {\em et al.}\ in Ref.~\cite{yin2014}: A spin system that undergoes a quantum phase transition at zero temperature and at some critical parameter value $\mu_c$, and for which quantum Kibble-Zurek scaling can be observed when ramping the parameter $\mu$ across the critical value, is elevated to an {\em open}\/ quantum system, i.e., coupled to an infinite bosonic bath. Performing now the same type of parameter ramp not at zero temperature, but at the nonzero temperature imposed by the bath, introduces an additional timescale into the problem, with the effect that the ``clean'' power-law scaling formerly observed at zero temperature gets polluted. To be precise, the relevant observable no longer exhibits power-law scaling at leading order in the ramp velocity. Instead, scaling behavior might only emerge approximately within a limited, possibly vanishing, parameter regime. While, even in this case, the critical exponents of the underlying equilibrium quantum phase transition still affect the nonequilibrium dynamics, it becomes practically impossible to read off critical exponents or perform a scaling analysis.

An approach to circumvent this problem, i.e., one that operates at nonzero temperatures and in the slow-ramp limit leads to clean scaling laws governed by quantum critical exponents, has been put forward in Ref.~\cite{KingKrielKastner23}. The key observation behind the ramping protocol proposed in that reference is that the nonzero temperatures, instead of being considered an obstacle, can be used to one's advantage: Clean scaling laws, governed by quantum critical exponents, can be obtained by ramping the temperature from a nonzero initial value to a nonzero finite value. In more technical terms, the main result of Ref.~\cite{KingKrielKastner23} is that, for temperature ramps $T(t)=T_i - vt$ with $t\in[0,t_f=T_i/v]$ in an open Kitaev chain at the quantum critical parameter value $\mu_c$, the total excitation density $\mathcal{E}$ satisfies the homogeneity relation
\begin{equation}\label{e:homogeneity}
\mathcal{E}\bigl(\ell^z T,\ell^z T_i,\ell^{-z(s+1)}\gamma/v\bigr) = \ell\, \mathcal{E}(T,T_i,\gamma/v)
\end{equation}
for arbitrary $\ell$. Here, $T_i$ and $T$ are the initial, respectively final, temperatures of the ramp, $v$ is the ramp velocity, and $\gamma$ is the coupling strength between system and bath. From the homogeneity relation \eqref{e:homogeneity}, various power-law scaling relations can be derived, which, like the homogeneity relation itself, depend on the exponent $s$ of the bath spectral density, as well as on the dynamical critical exponent $z$ of the underlying equilibrium quantum phase transition at $T=0$ and $\mu=\mu_c$. What is not featured in Eq.~\eqref{e:homogeneity}, however, is the quantum critical exponent $\nu$ which, in combination with $z$, quantifies the power-law divergence of the correlation length and is requisite for a complete characterization of the universality class of the equilibrium quantum phase transition.

\begin{figure}[t]
	\centering
	\includegraphics[width=\linewidth]{"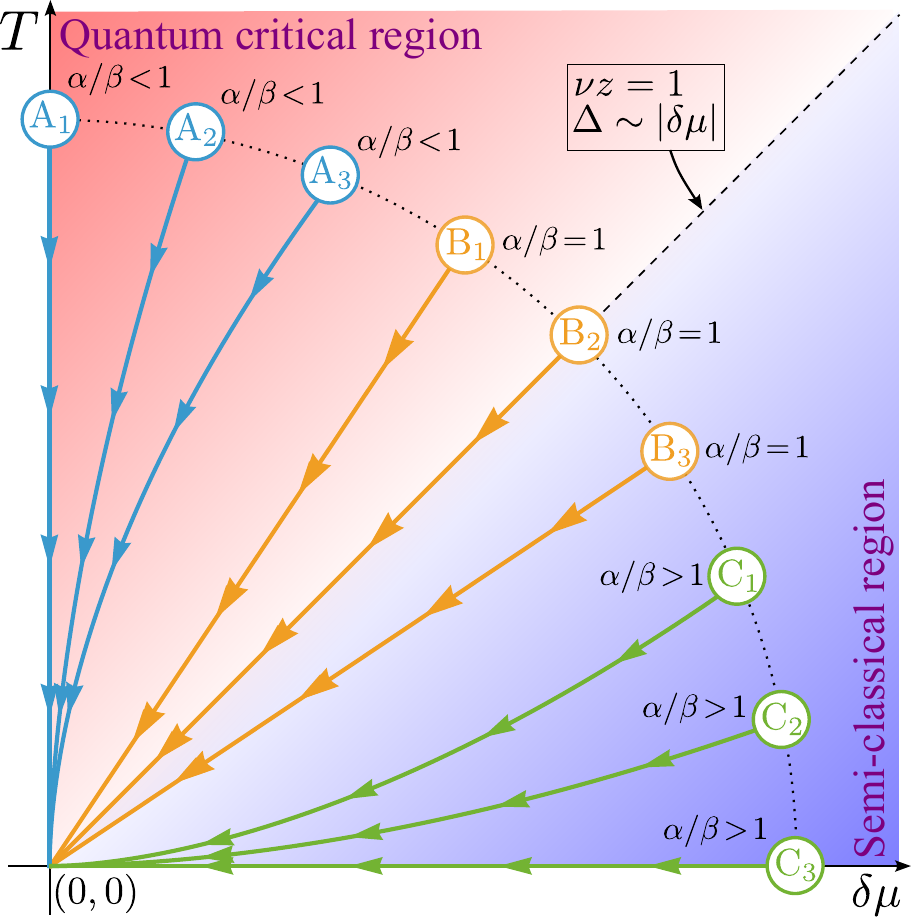"}
	\caption{A schematic representation of the $\dmu$--$T$ plane showing examples of the three classes of ramps described in Sec.~\ref{sec:Ramps}. Here we chose $\nuz=1$, as is the case for the Kitaev chain introduced in Sec.~\ref{sec:KitaevDefinition}.}
	\label{fig:ramp-paths}
\end{figure}

The main result of the present article is a Kibble-Zurek-type protocol that resolves this issue, i.e., that generates clean scaling laws that are governed by both the quantum critical exponents $z$ and $\nu$, while operating at nonzero temperatures. Hence, this protocol complements and generalizes the temperature ramps of Ref.~\cite{KingKrielKastner23} in such a way that the universality class of the underlying equilibrium quantum phase transition can be fully characterized on the basis of nonequilibrium data. This is achieved by considering protocols in which the temperature $T$ and the parameter $\mu$ are ramped simultaneously and in a way such that the quantum critical point is approached in a nonlinear fashion; see Fig.~\ref{fig:ramp-paths} for an illustration. We provide a complete solution of the asymptotic properties of such two-parameter ramps for arbitrary power-law ramps in the slow-ramp limit. By choosing suitable types of power-law ramps, both the coherent and incoherent parts of the open-system dynamics affect the excitation density $\mathcal{E}$ in a non-negligible way and lead to clean power-law scaling of the excitation density $\mathcal{E}$, involving the quantum critical exponents $z$ and $\nu$. These types of ramping protocols can be seen as successful and complete extensions of the quantum Kibble-Zurek protocol to finite temperatures. 

To guide the reader through the somewhat technical material of our work, we start in Sec.~\ref{sec:summaryofresults} with a brief nontechnical overview of our main results. In Sec.~\ref{sec:ModelDefinitionsandDynamics} we introduce an open-system version of spinless fermions on a one-dimensional lattice, with hopping and pairing terms, as well as thermalizing baths attached to each lattice site. This setup leads to dynamical equations for a set of expectation values which encode all the relevant dynamics of our system-bath model. Section \ref{sec:Ramps} is dedicated to introducing the three relevant classes of simultaneous $\mu$-$T$ ramps toward the critical point at $T=0$ and $\mu=\mu_c$. In Section \ref{sec:SlowRampDynamics} we derive suitable approximate forms for the relevant dynamical equations in the slow-ramp limit. This leads to the central scaling law for the system's excitation density at the end of the ramp. 

This groundwork culminates in the scaling analysis performed in Sec.~\ref{sec:ScalingAnalysis} where we derive expressions for the exponents that govern the leading-order scaling behavior of the excitation density in the slow-ramp limit. These exponents contain the critical exponents $\nu$ and $z$ associated with the quantum critical point. We find that the different ramp classes lead to different scaling behaviors, and in Sec.~\ref{sec:DiscussionResults} we benchmark these predictions against exact numerical results for the Kitaev chain with nearest-neighbor hopping and pairing terms. In  Sec.~\ref{sec:extraction} we discuss the use of these ramping protocols as a probe of (hypothetically unknown) critical exponents $\nu$ and $z$ from non-equilibrium dynamical data. We identify favorable protocols and parameter regimes for which obstruction of the leading-order asymptotic scaling behavior by subleading terms is relatively weak. An outlook on possible extensions and applications of our results is presented in Sec.~\ref{sec:SummaryConclusions}.

\section{Summary of results}
\label{sec:summaryofresults}
Sections~\ref{sec:ModelDefinitionsandDynamics}--\ref{sec:ScalingAnalysis} will focus on the technical details of the scaling analysis. It will be helpful to approach these technicalities with an overview of the final results already in mind, which we provide here. For the class of thermalizing open systems introduced in Sec.~\ref{sec:ModelDefinitionsandDynamics}, we consider ramping protocols where the bath temperature $T$ varies as 
\begin{equation}
T(t)=T_i(1-t/t_f)^\alpha
\end{equation}
from $T(0)=T_i$ down to $T(t_f)=0$. Here $t_f$ is the ramp duration, and $\alpha>0$ is a fixed parameter. Simultaneously, for the same ramp duration $t_f$, we additionally ramp a Hamiltonian parameter $\mu$ from $\mu(0)=\mu_i$ toward its critical value $\mu_c$ according to 
\begin{equation}
\delta\mu(t)=\delta\mu(0)(1-t/t_f)^\beta   
\end{equation}
with $\delta\mu(t)\equiv\mu(t)-\mu_c$ and fixed $\beta>0$. Together, these expressions define a ramp path in the $\dmu$-$T$ plane, starting at $(\dmu_i,T_i)$ and ending at the quantum critical point $(\dmu,T)=(0,0)$. The shape of this path is set by the ratio $\alpha/\beta$. By comparing this ratio to the critical exponent $\nu z$ that governs the closing of the excitation gap $\Delta\propto|\mu-\mu_c|^{\nu z}$, we identify three classes of ramps. These are denoted by class A ($\alpha/\beta<\nu z$), class B ($\alpha/\beta=\nu z$), and class C ($\alpha/\beta>\nu z$), and are shown schematically in Fig.~\ref{fig:ramp-paths} for the case where $\nu z=1$. As a measure of the nonadiabaticity of the dynamics, we consider the total excitation density $\mcE$ at the end of the ramp. For perfectly adiabatic dynamics, during which the system always remains in its instantaneous thermal equilibrium state, this excitation density should vanish at the end of the ramp, and a non-zero value of $\mcE(t_f)$ therefore signals a breakdown of adiabaticity. The dynamics of the excitation density is in principle determined through the master equation describing the open-system dynamics. However, it is found that the relevant physics can be captured through a reduced set of dynamical equations that describe the evolution of certain bilinear expectation values. 

Through an analysis of these dynamical equations we show that for slow ramps (large $t_f$) the excitation density obeys the scaling relation
\begin{multline}
	\mathcal{E}(t_f,\gamma,\kappa,\dmu_i,T_i)\\
	=a^{-1/z}\mathcal{E}(bt_f,\gamma/(a^s b),\kappa/(a b),a^{1/(\nuz)}\dmu_i,aT_i)
\end{multline}
for arbitrary $a,b>0$. Here, $\gamma$ sets the strength of the system--bath coupling and $\kappa$ is a bookkeeping parameter multiplying the unitary $i[\rho,H]$ term in the open-system master equation. Note that both the critical exponents $\nu$ and $z$ characterizing the quantum critical point appear explicitly in this scaling relation, as does the exponent $s$ associated with the bath spectrum. Through an appropriate choice of $a$ and $b$, tailored to the class of ramp under consideration, this scaling relation will allow us to extract the leading-order behavior of $\mcE(t_f)$ in the inverse ramp time $1/t_f$. This takes the form of a power-law relation
\begin{equation}\label{eq:scalingsummary}
    \mathcal{E}(t_f)=(t_0/t_f)^\zeta \mathcal{D},
\end{equation}
where the exponent $\zeta$ governs the vanishing of $\mcE(t_f)$ in the adiabatic limit where $t_f\rightarrow\infty$. Here $t_0$ is an arbitrary fixed time, and $\mathcal{D}$ a proportionality factor that is independent of $t_f$, but depends on the other ramp parameters. Consistent with the Kibble-Zurek framework for closed systems, we find that $\zeta$ encodes information about the equilibrium critical exponents $\nu$ and $z$, and now also of the bath spectral parameter $s$ as well as the powers $\alpha$ and $\beta$ characterizing the nonlinear ramp. For ramps in classes A and C we find the scaling powers
\begin{equation}\label{eq:zetasummary}
	  \zetaA =\frac{\alpha}{z(1+s\alpha)},\qquad\zetaC =\frac{\nu z\beta}{z(1+s\nu z\beta)},
\end{equation}
when $s\leq1$. For class B ramps, the powers $\alpha$ and $\beta$ satisfy $\alpha=\nu z\beta$, and $\zetaB$ is then given by either of the expressions in Eq.~\eqref{eq:zetasummary}. 
\begin{figure}[h!]
	\centering
	\includegraphics[width=\linewidth]{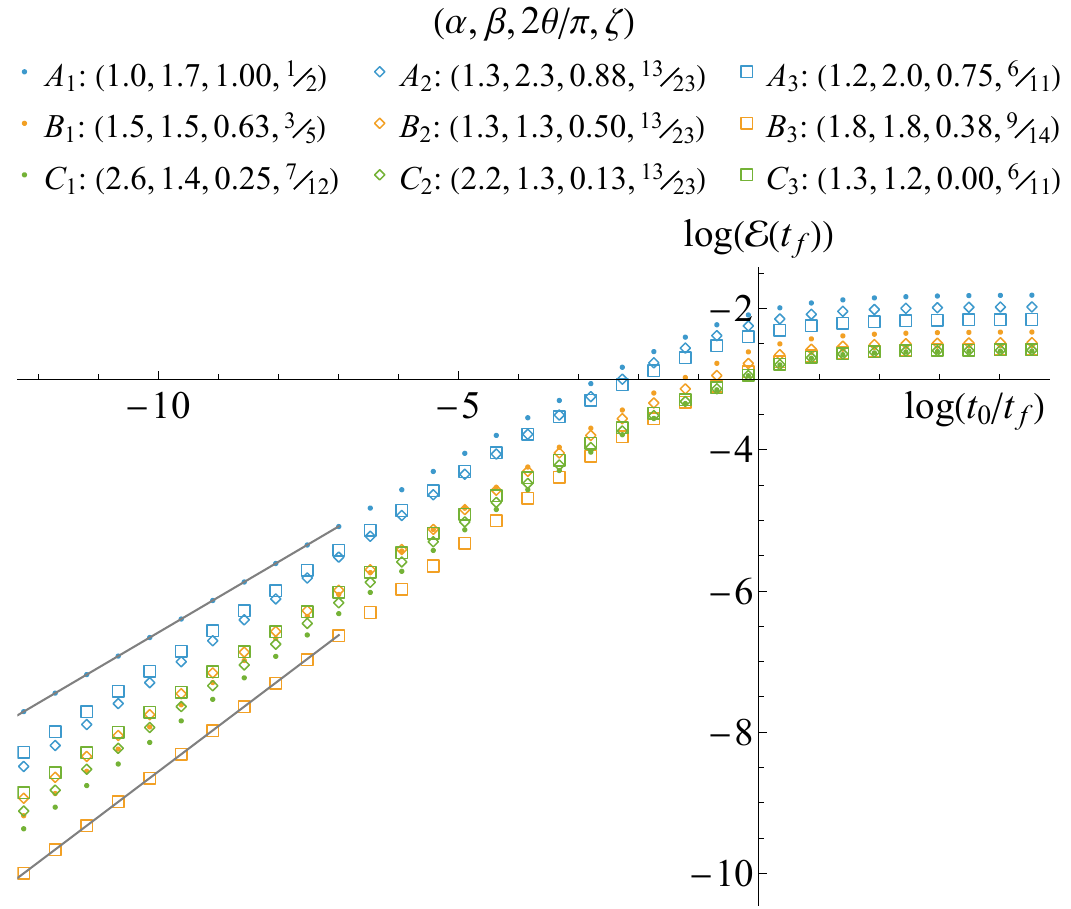}
    \includegraphics[width=\linewidth]{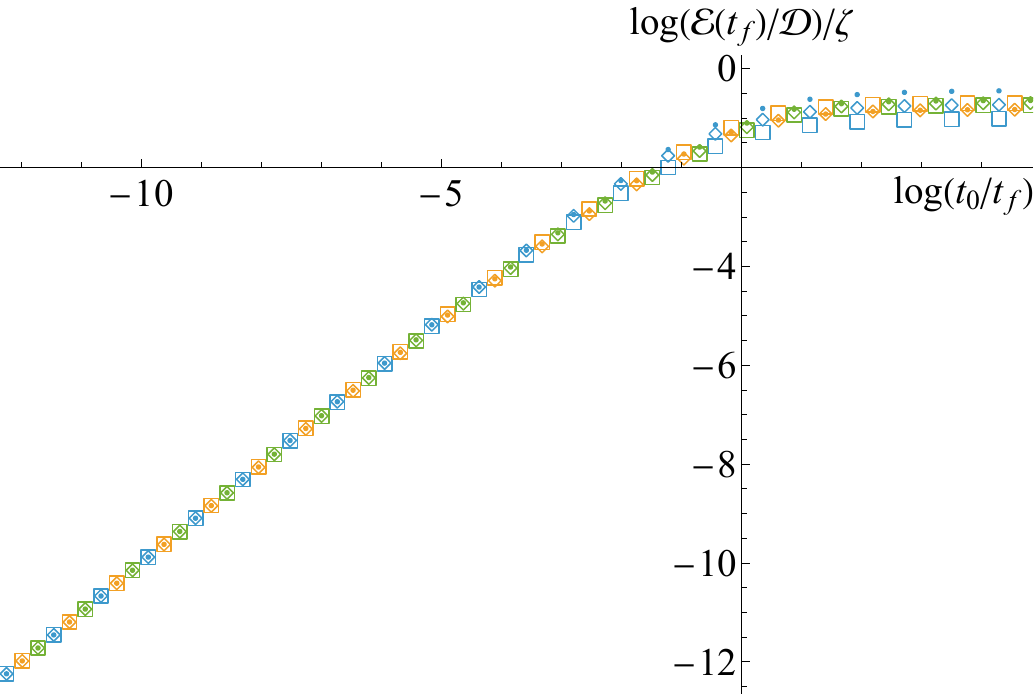}
	\caption{Top: Final excitation density $\mcE(t_f)$ for nine ramps, with ramp parameters as defined in the legend, obtained from exact numerical solutions of the dynamical equations \eqref{eq:FullDynamics} for the Kitaev model defined in Sec.~\ref{sec:KitaevDefinition}. The corresponding ramp paths are shown in Fig.~\ref{fig:ramp-paths}, and have starting points parametrized by $\theta$ as $(\dmu_i,T_i)=(-\cos(\theta),\sin(\theta))$. The linear behavior observed for slow ramps, i.e., for large $t_f$, reflects the scaling given schematically in Eq.~\eqref{eq:scalingsummary}, and in detail for the three ramp classes in Eqs.~\eqref{eq:ClassA_ED}, \eqref{eq:ClassB_ED} and \eqref{eq:ClassC_ED}. The predicted values of the scaling exponent $\zeta$ are given in the legend. For ramps $A_1$ and $B_2$, black lines with slopes corresponding to the predicted exponents $\zeta=1/2$ and $\zeta=9/14$ are shown for illustration. Bottom: An illustration of the data collapse resulting from the scaling relation in Eq.~\eqref{eq:scalingsummary} satisfied by $\mcE(t_f)$ for slow ramp velocities. Data for all nine ramps collapse to the same universal line, with unit slope, after an appropriate rescaling of $\mcE(t_f)$ and its logarithm.}
	\label{fig:collapse}
\end{figure}
These predictions are confirmed numerically for the Kitaev chain coupled to a thermal environment. Figure \ref{fig:collapse} (top) shows data for $\log(\mcE(t_f))$ as a function of $\log(t_0/t_f)$, obtained from an exact numerical solution of the dynamical equations for this model. Results for nine ramps across the three classes, with different starting points $(\dmu_i,T_i)$ and powers $\alpha$ and $\beta$, are shown. The ramp paths are those appearing in Fig.~\ref{fig:ramp-paths}. The straight-line behavior at large $t_f$ confirms the scaling in Eq.~\eqref{eq:scalingsummary}, with the slopes reflecting a range of distinct values of $\zeta$. See the caption for details. The scaling relation \eqref{eq:scalingsummary} also implies universality of the ratio $\mathcal{E}(t_f)/\mathcal{D}=(t_0/t_f)^\zeta$, in that its asymptotic behavior is determined by the exponent $\zeta$, but is insensitive to the precise starting point $(\dmu(0),T(0))$ of the ramp path. This results in the data collapse seen in Fig.~\ref{fig:collapse} (bottom), where $\log(\mcE(t_f)/\mathcal{D})/\zeta$ is shown as a function of $\log(t_0/t_f)$, with data points for all nine ramps lying on the same universal line.

\section{Model definition and dynamics}
\label{sec:ModelDefinitionsandDynamics}
We consider an open-system version of the Kitaev chain and generalizations thereof. The following properties make this family of models a ``minimal model'' for the physical situation we intend to study:
\begin{enumerate}
\setlength{\itemsep}{0pt}
\setlength{\parskip}{0pt}
\item The corresponding closed system undergoes an equilibrium quantum phase transition,\label{i:1}
\item it thermalizes under the open-system time evolution, and\label{i:2}
\item the open-system dynamics can be treated semi-analytically, and solved numerically for large system sizes.\label{i:3}
\end{enumerate}
See Ref.\ \cite{KingKastnerKriel} for details.

\subsection{System Hamiltonian}
\label{sec:SystemHamiltonian}
We consider a translationally invariant model of free spinless fermions moving on a chain with $L$ sites. The annihilation and creation operators associated with site $i$ are denoted by $c^\pdag_i$ and $c_i^\dag$, while those for a Fourier mode $k\in\{2\pi n/L:n\in\mathbb{Z}_L\}$ are $f^\pdag_k$ and $f^\dag_k$ with $f_k = L^{-1/2}\sum_{j}{e^{ijk}c_{j}}$. The system Hamiltonian in Fourier representation has the generic form 
\begin{equation}
	H = \frac{1}{2} \sum_{k}\psi_k^\dag \bm{H}_\text{BdG}(k) \psi^\pdag_k,
	\label{eq:SystemHamiltonianBdGForm}
\end{equation}
where $\psi_k = \rvec{f^\pdag_{k}}{f^\dag_{-k}}^{\!\mathsf{T}}$. Here,
\begin{equation}
	\bm{H}_\text{BdG}(k) = \mat{a_k & ib_k}{-ib_k & -a_k}
	\label{eq:BdGHamiltonian}
\end{equation}
is the $2 \times 2$ Bogoliubov--de Gennes Hamiltonian with entries satisfying $a_k=a_{-k}$ and $b_k=-b_{-k}$. The class of Hamiltonians \eqref{eq:SystemHamiltonianBdGForm} includes the well-known Kitaev chain \cite{Kitaev01} as well as its long-range extensions \cite{DuttaDutta17}. To diagonalize $H$ we introduce the Bogoliubov quasiparticle fermionic operators
\begin{equation}\label{eq:bogo}
    \eta^\pdag_k=\cos(\beta_k) f^\pdag_k-i\sin(\beta_k) f_k^\dag
\end{equation}
with Bogoliubov angle
\begin{equation}\label{eq:BogoliubovAngle}
	\beta_k=\frac{1}{2}\arctan\left(-\frac{b_k}{a_k}\right).
\end{equation}
In terms of these $\eta_k$-fermions the Hamiltonian reads
\begin{equation}
	H = \sum_{k}{\lambda_k \left(\eta_{k}^\dagger \eta_{k}^{\phantom\dag} -\tfrac{1}{2}\right)}
	\label{eq:HamiltonianDiagonalForm}
\end{equation}
with dispersion relation
\begin{equation}
    \lambda_k = \sqrt{a_k^2+b_k^2} = \lambda_{-k}.
\end{equation}

When investigating quantum phase transitions, we will be concerned with families of Hamiltonians $H(\mu)$, parametrized by a parameter $\mu$ through the matrix elements $a_k(\mu)$ and $b_k(\mu)$ in the Bogoliubov--de Gennes Hamiltonian \eqref{eq:BdGHamiltonian}. The occurrence of a second-order quantum phase transition is signaled by the vanishing of the excitation gap $\Delta(\mu)=\min_k \lambda_k(\mu)$ at a critical parameter value $\mu=\mu_c$ \cite{Kitaev01,DuttaDutta17}. 

Here we assume a family of Hamiltonians $H(\mu)$ that undergoes such a transition, and furthermore that it is the energy of the $k=0$ quasiparticle mode that vanishes at $\mu=\mu_c$. Close to that critical point the excitation gap then behaves as $\Delta=\lambda_0=c_2|\mu-\mu_c|^{\nuz}$, where $c_2$ is a constant and $\nu$ and $z$ are critical exponents. Exactly at the critical point the leading-order behavior of the dispersion relation is $\lambda_k=c_1|k|^z$, with $c_1$ another constant.

\subsection{Coupling to a thermal environment}
There are various ways of turning the closed system Hamiltonian of Sec.~\ref{sec:SystemHamiltonian} into a Markovian open quantum system that thermalizes to a Gibbs state with a given temperature $T$. One approach consists in postulating a Lindblad time evolution equation with jump operators that satisfy a detailed balance condition with respect to $T$ \cite{BacsiDora23}. Here we follow a different approach, which is more based on physical considerations and also leads to a thermalizing Lindblad dynamics. We consider the situation in which each lattice site is weakly coupled to one of $L$ identical and independent bosonic baths maintained at a temperature $T$. The overall quadratic structure of the system--bath Hamiltonian permits the formalism of D`Abbruzzo and Rossini \cite{DAbbruzzoRossini21} to be employed to derive a Markovian master equation for the system density matrix $\rho$. The result of this procedure is \cite{KingKastnerKriel}
\begin{align}
		\dot{\rho} = & -i\left[H,\rho\right]\nonumber\\
		& + \gamma\!\sum_{k,\sigma=\pm}\!\!\left(2 L_{k\sigma}^{\phantom{\dagger}}\rho L_{k\sigma}^\dagger - \left\{L_{k\sigma}^\dagger L_{k\sigma}^{\phantom{\dagger}},\rho\right\} \right),\label{eq:LindbladMasterEquation}
\end{align}
where the dimensionless parameter $\gamma$ sets the strength of the system--bath coupling. The jump operators are given by
\begin{equation}
	\quad L_{k,+} = \sqrt{\Gamma_{k,+}}\eta_{k}^\dagger,\qquad L_{k,-} = \sqrt{\Gamma_{k,-}}\eta^\pdag_{k},
\end{equation}
where 
\begin{subequations}
	\begin{align}
	\Gamma_{k,+} &= 2\pi\delta \lambda_k^s n_\text{\tiny BE}(\lambda_k/T),\\
	\Gamma_{k,-} &= 2\pi\delta \lambda_k^s (n_\text{\tiny BE}(\lambda_k/T)+1)
	\end{align}
\end{subequations}
contain the bath temperature $T$, and 
\begin{equation}
    n_\text{\tiny BE}(x)=\frac{1}{\exp(x)-1}
\end{equation}
is the Bose-Einstein distribution. Here $s$ and $\delta$ are positive parameters characterizing the bath spectral densities. The latter appears explicitly in the original full system--bath Hamiltonian from which the reduced Markovian master equation in Eq.~\eqref{eq:LindbladMasterEquation} is obtained \cite{DAbbruzzoRossini21,KingKastnerKriel}. When system parameters as well as bath parameters are kept fixed, this master equation will thermalize the system, in the sense that $\rho$ is converging to the Gibbs state $\rho^{\rm th}\sim\exp(-H/T)$ as $t\rightarrow\infty$ \cite{KingKastnerKriel}.

\subsection{Correlation function dynamics}
\label{sec:CorrelationDynamics}
Since the master equation is bilinear in the fermionic operators $\eta_k$, the method of third quantization \cite{Prosen08} can be used to calculate various quantities of interest. For the purpose of investigating Kibble-Zurek physics in open quantum systems, we consider the situation where both the system Hamiltonian $H$ and the bath temperature $T$ can be explicitly time-dependent, and our main quantity of interest is the mode occupation numbers $\ave{\eta_k^\dag \eta_k^{\phantom{\dag}}}$. Here the primary outcome of applying the third quantization approach is a matrix differential equation describing the dynamics of expectation values of bilinear fermionic operators. The translational invariance of both the system Hamiltonian and the bath configuration brings about a significant simplification. In particular, the matrix equation factorizes into a set of independent, lower-dimensional equations, one for each $\{k,-k\}$ pair of Fourier modes. Exploiting further relationships between these expectation values, the dynamics can be captured, for each value of $k$, in a set of coupled equations for $P_k=\ave{\eta_k^\dag \eta_k^{\phantom{\dag}}}$ and $C_k=\ave{\eta^\dag_k\eta^\dag_{-k}}$ \cite{KingKastnerKriel},
\begin{subequations}
	\label{eq:FullDynamics} 
	\begin{align}
	\deriv{P_k}{t}=&-R(\lambda_k,T)(P_k-P^{\text{th}}(\lambda_k/T))\nonumber\\
	&+2\deriv{\beta_k}{t} {\rm Im}(C_k),\label{eq:FullDynamicsA}\\
	\deriv{C_k}{t}=&-R(\lambda_k,T)C_k+2i\kappa\lambda_k C_k\nonumber\\
	&-2i\deriv{\beta_k}{t}(P_k-1/2).\label{eq:FullDynamicsB}
\end{align}
\end{subequations}
Here
\begin{equation}\label{eq:RDefinition}
	R(\lambda,T)=2\pi\gamma\delta\lambda^s \coth\left(\frac{\lambda}{2T}\right)
\end{equation} 
is the mode relaxation rate and
\begin{equation}
    P^{\text{th}}(x)=\frac{1}{\exp(x)+1}
\end{equation}
is the thermal equilibrium (Fermi-Dirac) distribution. We have introduced the dimensionless parameter $\kappa$ in the ${\sim}\lambda_k$ term of Eq.~\eqref{eq:FullDynamicsB}. This term originates from the $-i[H,\rho]$ term in the original master equation \eqref{eq:LindbladMasterEquation}. We will later use $\kappa$ as a bookkeeping device for tracking how this term is affected by certain parameter rescalings. 

Some remarks are in order here. First, note the appearance of derivatives of the Bogoliubov angle $\beta_k$ in Eq.~\eqref{eq:FullDynamics}. These derivatives emerge when the Hamiltonian parameter $\mu$ is varied in time; a scenario considered in the ramping protocols we introduce later. The derivative terms therefore result from working in the ``adiabatic frame'', i.e., in terms of the explicitly time-dependent $\eta_k$ fermions that diagonalize the Hamiltonian, rather than the ``diabatic frame" defined by the $f_k$ Fourier fermions. Secondly, note that when $\mu$, and therefore $\beta_k$, is kept fixed, the dynamics of $P_k$ decouples from that of $C_k$, and is described by a simple rate equation, which drives $P_k$ toward its thermal equilibrium value $P^{\rm th}(\lambda_k/T)$ at a rate set by $R(\lambda_k,T)$ \cite{KingKrielKastner23}. 

The equations in \eqref{eq:FullDynamics} capture all the relevant open-system dynamics of our model and contain the three coupled processes that govern the dynamics of the excitation probability $P_k$. We briefly discuss these here on a qualitative level. A more detailed discussion focused on ramp dynamics appears in Appendix~\ref{app:DynamicalProcesses}. As mentioned, the terms containing ${\rm d}\beta_k/{\rm d}t$ arise due to the explicit time-dependence of the Hamiltonian parameter $\mu$ and are responsible for generating excitations, i.e., increasing $P_k$ during the ramping protocols we will consider. The terms containing $\lambda_k$ account for the system's unitary dynamics and originate from the commutator in the master equation \eqref{eq:LindbladMasterEquation}. For slow variations of $\mu$ this term counteracts the generation of excitations due to the Hamiltonian driving and renders the unitary dynamics more adiabatic. Terms containing the relaxation rate $R(\lambda_k,T)$ describe the effect of the bath and steer the system toward thermal equilibrium. The full dynamics of $P_k$ is ultimately governed by an interplay of all three of these processes.

\section{Ramps toward the critical point}
\label{sec:Ramps}

The thermalizing master equation \eqref{eq:LindbladMasterEquation} permits the study of the nonequilibrium dynamics that results from time-dependent variations of the Hamiltonian parameter $\mu$ and the bath temperature $T$. At the level of the dynamical equations in \eqref{eq:FullDynamics}, $\mu$ enters through the mode energies and Bogoliubov angle, while $T$ resides in the relaxation rate and equilibrium distribution. The ability to vary both these parameters, either separately or simultaneously, allows a wide range of nonequilibrium ramping protocols to be explored. 

We consider ramps of the parameter $\mu$ and bath temperature $T$ that drive the system toward the quantum critical point at $\mu=\mu_c$ and $T=0$. Let $\delta\mu=|\mu-\mu_c|$ denote the distance of $\mu$ from its critical value. At $t=0$ the system is initialized in the thermal state corresponding to $T(0)=T_i$ and $\dmu(0)=\dmu_i$. This translates into the initial conditions $P_k(0)=P^{\rm th}(\lambda_k(\dmu_i)/T_i)$ and \mbox{$C_k(0)=0$} for Eq.~\eqref{eq:FullDynamics}. We then ramp $T$ and $\dmu$ to zero according to 
\begin{equation}\label{eq:RampDefinition}
	T(\tau)=T_i\tau^\alpha,\qquad\dmu(\tau)=\dmu_i\tau^\beta,
\end{equation}
with $\tau=1-t/t_f\in[0,1]$, where $t_f$ is the duration of the ramp. The positive powers $\alpha$ and $\beta$ determine the shape of the ramp path in the $\dmu$--$T$ plane. Unless $\alpha=\beta=1$, it is not possible to associate constant ramp velocities with these variations of $T$ and $\delta\mu$. Still, we can identify parameter combinations that fulfill essentially the same role. Writing \eqref{eq:RampDefinition} as
\begin{equation}\label{eq:RampDefinition2}
	\begin{split}
	T(\tau)&=T_i t_f^{-\alpha}(t_f-t)^\alpha,\\ \dmu(\tau)&=\dmu_i t_f^{-\beta}(t_f-t)^\beta,
	\end{split}
\end{equation}
we identify the prefactors
\begin{equation}\label{eq:RampSpeeds}
	v_T=T_i t_f^{-\alpha},\qquad v_{\mu}=\dmu_i t_f^{-\beta}.
\end{equation}
Together with $\alpha$ and $\beta$, these coefficients characterize the approach to the critical point. Abusing terminology slightly, we refer to $v_T$ and $v_\mu$ as the \emph{velocities} of the variations of $T$ and $\dmu$. For linear ramps with $\alpha=1$ and/or $\beta=1$, these quantities agree with the usual notion of a constant ramp velocity.

A ramp path in the $T$-$\dmu$ plane is described by the relationship
\begin{equation}
	\label{eq:CurveShape}
	\delta\mu_i^{\alpha/\beta}T=T­_i\delta\mu^{\alpha/\beta},
\end{equation}
which follows from combining the two expressions in Eq.~\eqref{eq:RampDefinition}. The ratio $\alpha/\beta$ therefore determines the shape of the ramp path in the $\dmu$-$T$ plane and leads to three classes of paths with, as we will see later, different excitation dynamics. The relevant comparison here is with the curve corresponding to the instantaneous minimum gap $\Delta\sim|\delta\mu|^{\nuz}$, the shape of which is set by the combination of critical exponents $\nuz$. Put differently, the three classes of ramps are identified through a comparison of the rates at which $T\sim|\delta\mu|^{\alpha/\beta}$ and $\Delta\sim|\delta\mu|^{\nuz}$ vanish as $\delta\mu\rightarrow0$. The classes are 
\begin{itemize}
	\item \textbf{Class A:} $T_i>0$, and $T$ vanishes more slowly than the instantaneous gap $\Delta$. This corresponds to a temperature-only ramp with $\delta\mu_i=0$, or to simultaneous $T$-$\mu$ ramps with $\alpha/\beta<\nuz$.
	\item \textbf{Class B:} $T$ vanishes at the same rate as the instantaneous gap $\Delta$. This occurs when $\alpha/\beta=\nuz$ and $T_i,\dmu_i>0$.
	\item \textbf{Class C:} $\delta\mu_i>0$, and $T$ vanishes more quickly than the instantaneous gap $\Delta$. This occurs for $\mu$-only ramps with $T_i=0$, or for simultaneous $T$-$\mu$ ramps with $\alpha/\beta>\nuz$.
\end{itemize}

Examples of ramp paths for these three classes are shown in Fig.~\ref{fig:ramp-paths} for the case where $\nuz=1$. Note that ramps with a starting point on the $T$-axis always fall in class A, while those starting on the $\dmu$-axis reside in class C. For ramps starting away from the two axes, the classification follows from the comparison of $\alpha/\beta$ with $\nu z$, with the precise initial point $(\dmu_i,T_i)$ not playing a role.

\section{Dynamics in the slow-ramp limit}
\label{sec:SlowRampDynamics}

\subsection{Background} 
\label{sec:SlowRampDynamicsBackground}
To quantify the breaking of adiabaticity that occurs in the course of the ramping protocols introduced in Sec.~\ref{sec:Ramps}, we will be monitoring the total excitation density \mbox{$\mathcal{E}(t)=\frac{1}{L}\sum_{k}P_k(t)$}. In the thermodynamic limit this quantity can be written as
\begin{equation}\label{eq:ExcitationDensity}
	\mathcal{E}(t)=\frac{1}{2\pi}\int_{-\pi}^{\pi}{\rm d}kP_k(t),
\end{equation}
where $P_k(t)$ is the solution of Eq.\ \eqref{eq:FullDynamics}. In the limit of slow ramps, i.e., for large values of the ramp time $t_f$, low-energy modes contribute dominantly to the final excitation density $\mathcal{E}(t_f)$, and it is their dynamics, close to the critical point, that is most relevant. This can be understood by noting that the relaxation rate $R(\lambda,T)$ in Eq.~\eqref{eq:RDefinition} is bounded from below by $2\pi\gamma\,\text{max}\{\lambda^s,(2T)^s\}$ when $s\leq1$, and is therefore finite for all modes except for $k=0$ precisely at the critical point. As we reduce the ramp velocity, i.e., increase $t_f$, the mode occupations $P_k$ therefore evolve more adiabatically, and can track the equilibrium occupation $P^{\rm th}(\lambda_k/T)$ even as the latter tends to zero. In the adiabatic limit where $t_f\rightarrow\infty$ all the mode occupations $P_{k\neq0}$ therefore vanish, and so does $\mcE(t_f)$.

For slow, finite-velocity ramps, the situation is quite different. As the critical point is approached, the relaxation rates of the low-energy modes decrease, preventing these modes from evolving adiabatically. In this sense, the system ``drops-out'' of equilibrium with the bath. Simultaneously, as $\dmu$ decreases and the excitation gap $\Delta$ closes, the variation of the system Hamiltonian results in the generation of additional excitations. The final excitation density $\mcE(t_f)$ is then a combination, though not simply a sum, of the residual excitations of the cooling process and the excess excitations generated by the Hamiltonian driving. In the following, we study properties of the final excitation density $\mcE(t_f)$ under simultaneous temperature and parameter ramps, with a particular focus on whether, at leading order in $1/t_f$, $\mcE(t_f)$ exhibits Kibble-Zurek-like power-law scaling.

\subsection{Local approximations}
As motivated in Sec.~\ref{sec:SlowRampDynamicsBackground}, it is the dynamics of the low-energy modes that, for slow ramps and close to the critical point, determines the final value of the excitation density $\mcE(t_f)$. With this in mind, we seek a simplified form of Eq.~\eqref{eq:FullDynamics} with which to describe the dynamics in this regime. This requires finding suitable approximations for the mode energy $\lambda_k$ and the Bogoliubov angle $\beta_k$ in a neighborhood of $(\delta\mu,k)=(0,0)$. We refer to these as \emph{local approximations}. It will be useful to make explicit the dependence of $\lambda_k$ and $\beta_k$ on $\dmu$, which we therefore denote by $\lambda(\dmu,k)$ and $\beta(\dmu,k)$. Recall from Sec.~\ref{sec:SystemHamiltonian} that the vanishing of the excitation gap at $\mu=\mu_c$ results in the leading-order power-law behavior $\lambda(0,k)=c_1|k|^z$ and $\lambda(\dmu,0)=c_2|\dmu|^{\nuz}$ of the dispersion relation. Any sensible local approximation of $\lambda(\dmu,k)$ must reproduce these relations. As a guide for how to proceed, we note that these scaling behaviors can be extracted through the limits $\lim_{q \rightarrow 0}\lambda(0,q^{1/z}k)/q=c_1|k|^z$ and $\lim_{q \rightarrow 0}\lambda(q^{1/(\nuz)}\dmu,0)/q=c_2|\dmu|^{\nuz}$. This suggests that the same limit procedure can produce a suitable local approximation of $\lambda(\dmu,k)$, valid also when both arguments are non-zero. We therefore define 
\begin{equation}\label{eq:LambdaApproximation}
	\Lambda(\dmu,k)=\lim_{q\rightarrow0}\frac{\lambda(q^{\frac{1}{\nuz}}\dmu,q^{\frac{1}{z}}k)}{q}
\end{equation}
to serve as a local approximation of the dispersion relation. In fact, this is equivalent to applying the same procedure to $a_k$ and $b_k$ individually, as these enter homogeneously in $\lambda(\dmu,k)=\sqrt{a_k^2+b_k^2}$. The analogous approximation for the Bogoliubov angle $\beta_k=(1/2)\arctan(-b_k/a_k)$ is 
\begin{equation}\label{eq:BogoliubovApproximation}
	\mathcal{B}(\dmu,k)=\lim_{q\rightarrow0}\beta(q^{\frac{1}{\nuz}}\dmu,q^{\frac{1}{z}}k).
\end{equation}
Note that $\Lambda$ and $\mathcal{B}$ are defined for all values of $\dmu\geq0$ and $k\in\mathbb{R}$, and that they satisfy the scaling identities
\begin{subequations}
\begin{align}\label{eq:ScalingRelations1}
	\Lambda(a^{\frac{1}{\nuz}}\dmu,a^{\frac{1}{z}}k)&=a\Lambda(\dmu,k),\\
	\mathcal{B}(a^{\frac{1}{\nuz}}\dmu,a^{\frac{1}{z}}k)&=\mathcal{B}(\dmu,k),
\end{align}
\end{subequations}
for all $a>0$. (The arbitrary scaling factor $a$ should not be confused with the matrix element $a_k$ from Eq.~\eqref{eq:BdGHamiltonian}.)  It will be useful to label the $k\geq0$ modes using their energies at the critical point. This amounts to swapping $k$ for $\epsilon$, where the two are related by $\epsilon=\Lambda(0,k)=c_1|k|^z$. Since $\Lambda(0,k)=\Lambda(0,-k)$, the $k<0$ modes can later be accounted for by a factor of two in the density of states. With $\Lambda$ and $\mathcal{B}$ considered as functions of $\epsilon$, the scaling relations above become
\begin{subequations}
\begin{align}
	\Lambda(a^{\frac{1}{\nuz}}\dmu,a\epsilon)&=a\Lambda(\dmu,\epsilon),\label{eq:ScalingRelations2a}\\
	\mathcal{B}(a^{\frac{1}{\nuz}}\dmu,a\epsilon)&=\mathcal{B}(\dmu,\epsilon).\label{eq:ScalingRelations2b}
\end{align}
\end{subequations}
Returning to Eq.~\eqref{eq:FullDynamics}, we now replace $\lambda$ and $\beta$ by the local approximations $\Lambda$ and $\mathcal{B}$ to obtain
\begin{subequations}
	\label{eq:FullDynamicsLocal}
	\begin{align}
	\deriv{P}{t}&=-R(\Lambda,T)(P-P^{\rm th}(\Lambda/T))+2\deriv{\mathcal{B}}{t} {\rm Im}(C),\\
	\deriv{C}{t}&=-R(\Lambda,T)C+2i\kappa\Lambda C-2i\deriv{\mathcal{B}}{t}(P-1/2),
\end{align}
\end{subequations}
and the initial values for $P$ and $C$ become $P(0)=P^{\rm th}(\Lambda(\dmu_i,\epsilon)/T_i)$ and $C(0)=0$. These time evolution equations, together with the local approximations defined in Eqs.~\eqref{eq:LambdaApproximation} and \eqref{eq:BogoliubovApproximation}, capture the dynamics of the low-energy modes asymptotically for slow ramps close to the critical point, which is what we set out to do. At the same time, because local approximations have simpler functional forms than their original counterparts, Eq.~\eqref{eq:FullDynamicsLocal} is more amenable to a scaling analysis and will be the basis of our derivation of scaling laws in the remainder of the article.

The final excitation density $\mathcal{E}=\mathcal{E}(t_f)$ in Eq.~\eqref{eq:ExcitationDensity} can now be written as
\begin{align}\label{eq:ExcitationDensityEnergyIntegralLocal}
	\mathcal{E}&=\int_0^\infty{\rm d}\epsilon\tilde{\rho}(\epsilon)P(\epsilon,t_f,\gamma,\kappa,\dmu_i,T_i),
\end{align}
where the density of states within the local approximation is
\begin{equation}\label{eq:RhoLocal}
	\tilde{\rho}(\epsilon)=\frac{1}{\pi z c_1^{1/z}}\epsilon^{1/z-1}.
\end{equation} 
Our focus on the slow-ramp limit allowed the upper bound of the integral to be extended to infinity, since only the low-energy modes will contribute non-negligibly to the final excitation density. 

\subsection{Example: Kitaev chain Hamiltonian}
\label{sec:KitaevDefinition}
To illustrate the working of the local approximation in Eqs.~\eqref{eq:LambdaApproximation} and \eqref{eq:BogoliubovApproximation} it is useful to consider the Kitaev chain \cite{Kitaev01} as a concrete example. This is also the model we will later use for performing comparisons to numerical results. The Kitaev chain Hamiltonian is a special case of Eqs.~\eqref{eq:SystemHamiltonianBdGForm} and \eqref{eq:BdGHamiltonian}, defined by
\begin{equation}
	a_k=2(\mu+J\cos(k)),\qquad b_k=\Delta_p \sin(k).
	\label{eq:Kitaevabdef}
\end{equation}
Here $J$ and $\Delta_p$ are respectively hopping and pairing strengths. The critical value of $\mu$ is $\mu_c=-J$, while the critical exponents are $\nu=z=1$. The mode energy $\lambda_k$ and Bogoliubov angle $\beta_k$ from Sec.~\ref{sec:SystemHamiltonian} now have the explicit forms
\begin{subequations}
\label{eq:KitaevLambdaBeta}
\begin{align}
    \lambda(\dmu,k)&=\sqrt{4(\delta\mu-J+J \cos(k))^2+\Delta_p^2 \sin(k)^2},\\
    \beta(\dmu,k)&=-\frac{1}{2}\arctan\left[\frac{\Delta_p \sin(k)}{2(\delta\mu-J+J\cos(k)}\right].
\end{align}
\end{subequations}
Applying the definitions in Eqs.~\eqref{eq:LambdaApproximation} and \eqref{eq:BogoliubovApproximation} yields the corresponding local approximations
\begin{subequations}
\label{eq:KitaevLocalApproximations}
\begin{align}
	\Lambda(\dmu,k)&=\sqrt{4\dmu^2+k^2\Delta_p^2}\,,\\
	\mathcal{B}(\dmu,k)&=\frac{1}{2}\arctan\left(-\frac{\Delta_p k}{2\dmu}\right),
\end{align}
\end{subequations}
while $\epsilon=\Lambda(0,k)=|\Delta_p k|$.

\subsection{Fixed velocity ramps}
\label{sec:ConstandSpeedRamps}
The ramps we consider start from the initial values $(\dmu_i,T_i)$ and then approach  $(\dmu,T)=(0,0)$ over a time interval of $t_f$. In the limit of slow ramps, i.e., for large $t_f$, the ramp velocities $v_T$ and $v_\mu$ in Eq.~\eqref{eq:RampSpeeds} tend to zero. However, the analysis in Sec.~\ref{sec:ScalingAnalysis} will lead us to also consider ramps taking place at fixed velocities over an increasingly long time interval. We pause here to introduce some notation for this case. Now $t_f$, $v_T$, and $v_\mu$ are the natural input parameters, with the ramp's starting point $(\dmu_i,T_i)=(v_\mu t_f^{\beta},v_T t_f^{\alpha})$ determined in terms of these. In particular, increasing $t_f$ extends the ramp path along the curve described by Eq.~\eqref{eq:CurveShape} in order to keep the two velocities fixed. The excitation density is now naturally a function of $t_f$, $v_T$, and $v_\mu$, namely
\begin{gather}
	\begin{aligned}
		&\mathcal{E}(t_f,\gamma,\kappa,v_\mu,v_T)\\
		&=\int_0^\infty{\rm d}\epsilon\tilde{\rho}(\epsilon)P(\epsilon,t_f,\gamma,\kappa,\dmu_i=v_\mu t_f^{\beta},T_i=v_T t_f^{\alpha}).
	\end{aligned}
\end{gather}
We are interested in the $t_f\rightarrow\infty$ limit of this excitation density, denoted as
\begin{equation}
    \label{eq:DDefinition}
	\mathcal{D}(\gamma,\kappa,v_\mu,v_T)=\lim_{t_f\rightarrow\infty}\mathcal{E}(t_f,\gamma,\kappa,v_\mu,v_T).
\end{equation}
The existence of this limit follows from the same reasoning as was used in Sec.~\ref{sec:SlowRampDynamicsBackground} to motivate the local form of the dynamical equations in Eq.~\eqref{eq:FullDynamicsLocal}. In short, by increasing $t_f$ with the ramp velocities kept fixed, the ramp path is extended further away from the critical point at $(\dmu,T)=(0,0)$, into regions of the $\dmu$-$T$ plane where the mode relaxation rates $R(\lambda,T)\geq2\pi\gamma\text{max}\{\lambda^s,(2T)^s\}\geq2\pi\gamma\text{max}\{\Delta^s,(2T)^s\}$ are large. In these regions, the system evolves adiabatically, remaining in equilibrium with the bath. Extending the ramp path in this way, therefore, has no impact on the final value of the excitation density, which results entirely from the breakdown of adiabaticity close to the critical point. This implies that the excitation density $\mcE$ on the right-hand side of Eq.~\eqref{eq:DDefinition} will tend to a finite value as $t_f$ increases.

\subsection{Central scaling identity}
In this section we show that the locally approximated dynamical equations \eqref{eq:FullDynamicsLocal} lead to a scaling identity for the excitation density. Based on this identity, scaling laws will be derived in Sec.~\ref{sec:ScalingAnalysis}. Note that the explicit time dependence on the right-hand side of Eq.~\eqref{eq:FullDynamicsLocal} enters only through the dimensionless combination \mbox{$\tau=1-t/t_f\in[0,1]$}, specifically via the $\dmu(\tau)$ and $T(\tau)$ dependencies of $\Lambda$, $\mathcal{B}$, and $R(\Lambda,T)$. Trading $t$ for $\tau$ in Eq.~\eqref{eq:FullDynamicsLocal} yields the set of equations
\begin{subequations}\label{eq:FullDynamicsLocalDimensionless}
	\begin{align}
	-\dot{P}&=-t_f R(\Lambda,T)(P-P^{\rm th}(\Lambda/T))+2\dot{\mathcal{B}} {\rm Im}(C),\label{eq:FullDynamicsLocalDimensionless1}\\
	-\dot{C}&=-t_f R(\Lambda,T)C+2it_f\kappa\Lambda C-2i\dot{\mathcal{B}}(P-1/2),\label{eq:FullDynamicsLocalDimensionless2}
\end{align}
\end{subequations}
where the dot denotes a derivative with respect to $\tau$. We can identify four relevant parameter combinations entering on the right-hand sides of Eqs.\ \eqref{eq:FullDynamicsLocalDimensionless1} and \eqref{eq:FullDynamicsLocalDimensionless2}, and also in the initial condition $P(0)=P^{\rm th}(\Lambda(\dmu_i,\epsilon)/T_i)$, namely
\begin{subequations}\label{eq:ParameterCombinations}
\begin{align}
	&t_f\gamma\Lambda^s(\dmu,\epsilon),&&\Lambda(\dmu,\epsilon)/T,\\
	&t_f\kappa\Lambda(\dmu,\epsilon),&&\mathcal{B}(\dmu,\epsilon),
\end{align}
\end{subequations}
the first of which forms part of $R(\Lambda,T)$, as in Eq.~\eqref{eq:RDefinition}. The definitions of $\dmu$ and $T$ in Eq.~\eqref{eq:RampDefinition}, combined with the scaling properties of $\Lambda$ and $\mathcal{B}$ in Eqs.~\eqref{eq:ScalingRelations2a} and \eqref{eq:ScalingRelations2b}, imply that all four quantities in Eq.\ \eqref{eq:ParameterCombinations} are invariant under the rescaling 
\begin{align}
	&\!\!\!(\epsilon,t_f,\gamma,\kappa,\dmu_i,T_i)\nonumber\\
	&\!\!\!\rightarrow(a\epsilon,bt_f,\gamma/(a^s b),\kappa/(a b),a^{1/(\nuz)}\dmu_i,aT_i),\label{eq:Rescaling}
\end{align}
for all $a,b>0$. It follows that the dynamical equations in Eq.~\eqref{eq:FullDynamicsLocalDimensionless} and their initial conditions are invariant under this rescaling, and the same holds for the excitation probability $P$ at the end of the ramp. In turn, this implies a scaling relation for the excitation density $\mathcal{E}$ in Eq.~\eqref{eq:ExcitationDensityEnergyIntegralLocal}. Rescaling the arguments of $P$ in Eq.~\eqref{eq:ExcitationDensityEnergyIntegralLocal} according to Eq.~\eqref{eq:Rescaling}, and then rescaling the integration variable as $\epsilon\rightarrow\epsilon/a$ leads to the equality
\begin{align}
	&\mathcal{E}(t_f,\gamma,\kappa,\dmu_i,T_i)\nonumber\\
	&=a^{-1/z}\mathcal{E}(bt_f,\gamma/(a^s b),\kappa/(a b),a^{1/(\nuz)}\dmu_i,aT_i),\label{eq:ExcitationDensityScaling1}
\end{align}
where the form of $\tilde{\rho}(\epsilon)$ from Eq.~\eqref{eq:RhoLocal} was used. Equation~\eqref{eq:ExcitationDensityScaling1} can be viewed as a two-parameter generalization of a generalized homogeneous function. At this stage, the two parameters $a$ and $b$ are still arbitrary. Our goal, as will be explained in Sec.~\ref{sec:ScalingAnalysis}, is to choose $a$ and $b$ such as to isolate the leading-order dependence of $\mathcal{E}$ on $t_f$ within the $a^{-1/z}$ pre-factor on the right of Eq.~\eqref{eq:ExcitationDensityScaling1}. This suggests choosing $a$ and $b$ as powers of $1/t_f$. Specifically, we set
\begin{equation}
	\label{eq:abfDefinition}
	a=f^{p/s},\qquad b=f^{-r}
\end{equation}
with
\begin{equation}
	\label{eq:abfDefinition2}
	f=t_f/t_0,
\end{equation}
where $t_0$ is an arbitrary fixed time, introduced to render $a$ and $b$ dimensionless. The powers $p$ and $r$, at present arbitrary, will later be fixed based on specifics of the ramp under consideration and the value of the bath spectral parameter $s$ in Eq.~\eqref{eq:RDefinition}. For $a$ and $b$ as in Eq.~\eqref{eq:abfDefinition}, the scaling identity \eqref{eq:ExcitationDensityScaling1} becomes
\begin{align}\label{eq:Homogeneity}
		&\mathcal{E}(t_f,\gamma,\kappa,\dmu_i,T_i)=f^{-p/(sz)}\nonumber\\
		&\times\mathcal{E}(t_0 f^{1-r},f^{r-p}\gamma,f^{r-p/s}\kappa,f^{p/(s\nuz)}\dmu_i,f^{p/s}T_i).
\end{align}
We abbreviate this result in the form
\begin{equation}\label{eq:ExcitationDensityScaling2}
	\mathcal{E}=f^{-\zeta}\bmcE\qquad{\rm with}\qquad\zeta=\frac{p}{sz},
\end{equation}
where 
\begin{equation}
	\mcE=\mathcal{E}(t_f,\gamma,\kappa,\dmu_i,T_i),\quad\bmcE=\mathcal{E}(\bar{t}_f,\bar{\gamma},\bar{\kappa},\widebar{\dmu}_i,\widebar{T}_i),
\end{equation}
and  
\begin{align}
	&(\bar{t}_f,\bar{\gamma},\bar{\kappa},\widebar{\dmu}_i,\widebar{T}_i)\nonumber\\
	&=(t_0 f^{1-r},f^{r-p}\gamma,f^{r-p/s}\kappa,f^{p/(s\nuz)}\dmu_i,f^{p/s}T_i).\label{eq:RescaledParameters}
\end{align}
By choosing the parameters $a$ and $b$ according to Eq.\ \eqref{eq:abfDefinition} and thereby obtaining the relation \eqref{eq:Homogeneity}, we established that the total excitation density $\mathcal{E}$ is a generalized homogeneous function with $f$ as a scaling parameter. Such a homogeneity relation implies the existence of scaling relations between various of the quantities involved, as we shall see in the following. Equation \eqref{eq:ExcitationDensityScaling2} expresses a quantitative relationship between the final excitation densities $\mcE$ and $\bmcE$ of two ramps with different parameter values. The former is that of the physical ramp under consideration, characterized by the original set of parameters $(t_f,\gamma,\kappa,\dmu_i,T_i)$. The latter is the excitation density of an auxiliary ramp, defined by the rescaled set of parameters $(\bar{t}_f,\bar{\gamma},\bar{\kappa},\widebar{\dmu}_i,\widebar{T}_i)$. The ramp velocities for the auxiliary ramps follow from Eq.~\eqref{eq:RampSpeeds} as
\begin{subequations}\label{eq:RampsSpeeds2}
\begin{align}
	\bar{v}_T&=\widebar{T}_i\bar{t}_f^{-\alpha}=f^{p/s-\alpha(1-r)}T_it_0^{-\alpha},\\
	\bar{v}_\mu&=\widebar{\dmu}_i\bar{t}_f^{-\beta}=f^{p/(s\nuz)-\beta(1-r)}\dmu_it_0^{-\beta}.
\end{align}
\end{subequations}

\section{Scaling analysis}
\label{sec:ScalingAnalysis}

\subsection{Overview of strategy}
\label{sec:StrategyOverview}
Our goal is to use Eq.~\eqref{eq:Homogeneity} to extract the scaling behavior of $\mcE$ with respect to $1/t_f$ in the slow-ramp limit. To this end, we must choose the powers $p$ and $r$ such that the $f^{-\zeta}=(t_0/t_f)^\zeta$ prefactor in Eq.~\eqref{eq:ExcitationDensityScaling2} captures the leading-order dependence of $\mcE$ on $1/t_f$. This requires that the excitation density $\bmcE$ of the auxiliary ramp has a finite $f,t_f\rightarrow\infty$ limit, in which case it will hold to leading order in $1/t_f$ that
\begin{equation}
	\label{eq:ExcitationDensityScaling3}
	\mcE=(t_0/t_f)^{-\zeta}\lim_{f,t_f\rightarrow\infty}\bmcE.
\end{equation}
The $\zeta=p/(sz)$ power therefore captures the power-law scaling of $\mcE$ with the inverse ramp time, while the limiting value of $\bmcE$ acts as a proportionality constant. To achieve these finite limit values, specific choices must be made for $p$ and $r$. The reasoning that leads us to suitable choices is as follows. First, note that both $\mcE$ and $\zeta$ are independent of $r$, so the same is true of $\bmcE$. Since this allows $r$ to be chosen freely, we make a choice that simplifies the $f$ dependence of the arguments of $\bmcE$, as listed in Eq.~\eqref{eq:RescaledParameters}. Specifically, we choose $r$ such that $\bar{\kappa}=f^{r-p/s}\kappa$ or $\bar{\gamma}=f^{r-p}\gamma$ is independent of $f$, and therefore unaffected by the $f\rightarrow\infty$ limit. This identifies $r=p$ and $r=p/s$ as convenient choices. What remains is to choose $p$. This choice will also fix the powers of $f$ in the expressions for $\bar{v}_T$ and $\bar{v}_\mu$ in Eq.~\eqref{eq:RampsSpeeds2}. Since we want $\lim_{f\rightarrow\infty}\bmcE$ to be finite, we should avoid a scenario where $\bar{v}_T$ or $\bar{v}_\mu$ diverges, or where both these velocities tend to zero. This suggests choosing $p$ such that at least one of these velocities remains fixed and independent of $f$, while the other, if not also fixed, tends to zero. The first of these conditions identifies $p=s\alpha(1-r)$ and $p=s\nu z\beta(1-r)$ as appropriate choices. These two options for $r$ and two for $p$ lead to four ``rescaling schemes'', the first two of which are described in the next section. The choice of which particular scheme to use is ultimately determined by the details of the ramp under consideration, as will be seen in Sec.~\ref{sec:ScalingLaws}.

\subsection{Rescaling schemes}
\label{sec:RescalingSchemes}
The strategy outlined in Sec.~\ref{sec:StrategyOverview} leads to four choices for the powers $r$ and $p$ in the scaling identity \eqref{eq:Homogeneity}. We describe the first two of these below, with the remaining two outlined in Appendix \ref{app:RescalingSchemes}. 
\begin{itemize}[leftmargin=*]
	\item \textbf{Scheme 1:} We require that $\bar{\gamma}$ and $\bar{v}_T$ are independent of $f$. Since $\bar{\gamma}=f^{r-p}\gamma$, we choose $r=p$. Then, from $\bar{v}_T=f^{p/s-\alpha(1-r)}T_it_0^{-\alpha}$ it follows that 
    \begin{equation}
		\label{eq:Scheme1_rp}
		r=p=\frac{s\alpha}{1+s\alpha}
	\end{equation}
    will render $\bar{v}_T$ independent of $f$. Defining \mbox{$q=f^{1-r}$}, the parameters of $\bmcE$ can be expressed as
	\begin{align}
		&(\bar{t}_f,\bar{\gamma},\bar{\kappa},\widebar{\dmu}_i,\widebar{T}_i)\nonumber\\
		&=(qt_0,\gamma,q^{\alpha(s-1)}\kappa,q^{\alpha/(\nuz)}\dmu_i,q^\alpha T_i),\label{eq:Scheme1_parameters}
	\end{align}
	with the ramp velocities becoming
	\begin{equation}
		\label{eq:Scheme1_speeds}
		\bar{v}_T=t_0^{-\alpha}T_i,\qquad
        \bar{v}_\mu=q^{\alpha/(\nuz)-\beta}t_0^{-\beta}\dmu_i.
	\end{equation}
	
    \item \textbf{Scheme 2:} Requiring $\bar{\gamma}$ and $\bar{v}_\mu$ to be independent of $f$ leads to the choice
	\begin{equation}
		\label{eq:Scheme2_rp}
		r=p=\frac{s\nuz\beta}{1+s\nuz\beta}.
	\end{equation}
	Setting $q=f^{1-r}$, the parameters of $\bmcE$ are
	\begin{align}
		&(\bar{t}_f,\bar{\gamma},\bar{\kappa},\widebar{\dmu}_i,\widebar{T}_i)\nonumber\\
		&=(qt_0,\gamma,q^{\nuz\beta(s-1)}\kappa,q^\beta\dmu_i,q^{\nuz\beta} T_i),\label{eq:Scheme2_parameters}
	\end{align}
	with the ramp velocities becoming
	\begin{equation}
		\label{eq:Scheme2_speeds}
		\bar{v}_T=q^{\nuz\beta-\alpha}t_0^{-\alpha}T_i,\qquad \bar{v}_\mu=t_0^{-\beta}\dmu_i.
	\end{equation}
\end{itemize}

\subsection{Scaling laws for $\mu$-$T$ ramps}
\label{sec:ScalingLaws}
We proceed to show how combining the homogeneous function property \eqref{eq:ExcitationDensityScaling2} with the appropriate rescaling scheme allows the value of the scaling exponent $\zeta$ to be extracted. As was outlined in Sec.~\ref{sec:StrategyOverview}, the guiding principle for matching a particular ramp class with a rescaling scheme is to ensure that the ramp velocities $\bar{v}_\mu$ and $\bar{v}_T$ in Eq.~\eqref{eq:RampsSpeeds2}, which are associated with $\bmcE$, remain bounded in the $f,t_f\rightarrow\infty$ limit. Deriving such scaling laws in full generality, there are several different cases to consider. This includes classes A, B, and C of ramp types we classified in Sec.~\ref{sec:Ramps}; the cases of sub-ohmic ($s<1$), ohmic ($s=1$), and super-ohmic ($s>1$) baths; as well as the special case of purely unitary dynamics with $\gamma=0$. The unitary case is best treated separately and is presented in Appendix~\ref{app:CoherentScaling}. The super-ohmic case introduces additional technical complications, is not nearly as straightforward to deal with as the other cases, and will not be presented here. In short, we find for the super-ohmic case the same scaling laws as for the ohmic and sub-ohmic cases for ramp classes A and B. For class C, the scaling is different, and reduces to that of a ramp governed by purely unitary dynamics, as treated in Appendix \ref{app:RescalingSchemes}.

The remaining cases, i.e., the ramp classes A, B, and C for the sub-ohmic and ohmic cases, respectively, will be discussed in the following. We first formulate our results as scaling laws with respect to the inverse ramp time $1/t_f$, and then show in Sec.~\ref{sec:ScalingWithRampVelocities} how to rephrase these as scaling laws with respect to the ramp velocities.

\subsubsection{Ramps in class A}
\label{sec:RampsA}
This class comprises temperature-only ramps with $\delta\mu_i=0$, as well as simultaneous $T$-$\mu$ ramps for which $\alpha/\beta<\nuz$. We choose Scheme 1, setting $r$ and $p$ as in \eqref{eq:Scheme1_rp}. From Eqs.~\eqref{eq:Scheme1_parameters} and \eqref{eq:Scheme1_speeds} it follows that 
\begin{align}
	\lim_{f,q\rightarrow\infty}\bar{\kappa}&=\bar{\kappa}_\infty\equiv\begin{cases} \kappa & {\rm if}\ s=1\\0 & {\rm if}\ s<1 \end{cases},\label{eq:KappaBarLimit}\\
	\lim_{q\rightarrow\infty}\bar{v}_\mu&=0,
\end{align}
where we introduced the shorthand $\bar{\kappa}_\infty$. Combined with the scaling identity \eqref{eq:ExcitationDensityScaling3} this implies that, to leading order in $1/t_f$, we have
\begin{align}
	&\mathcal{E}(t_f,\gamma,\kappa,\dmu_i,T_i)\nonumber\\
	&=(t_0/t_f)^\zetaA\mathcal{D}(\gamma,\bar{\kappa}_\infty,\bar{v}_\mu=0,\bar{v}_T=t_0^{-\alpha}T_i)\label{eq:ClassA_ED}
\end{align}
with
\begin{equation}
	\label{eq:ClassA_ScalingPower}
    \zetaA\equiv\frac{\alpha}{z(1+s\alpha)}.
\end{equation}
The excitation density $\mathcal{D}$ on the right-hand side of Eq.~\eqref{eq:ClassA_ED} is that of an infinite-time, fixed velocity ramp, as defined in Eq.~\eqref{eq:DDefinition}. Since the velocity associated with the variation of $\dmu$ is zero, this is a temperature-only ramp. In fact, we note that the dependence on $\dmu_i$ has also been eliminated in \eqref{eq:ClassA_ED}. To leading order in $1/t_f$, all ramps in class A are therefore equivalent to temperature-only ramps. The scaling power identified in Eq.~\eqref{eq:ClassA_ScalingPower} also matches that found in Ref.~\cite{KingKrielKastner23} for such ramps, obtained by solving the rate equation for the excitation probability explicitly. 

\subsubsection{Ramps in class B}
Class B consists of simultaneous $T$-$\mu$ ramps for which $T_i,\dmu_i>0$ and $\alpha/\beta=\nu z$. Here, Schemes 1 and 2 are equivalent, i.e., lead to identical choices of $r$ and $p$. We set $r$ and $p$ as in \eqref{eq:Scheme1_rp}, and adopt the $\bar{\kappa}_\infty$ shorthand from Eq.~\eqref{eq:KappaBarLimit}. From Eq.~\eqref{eq:ExcitationDensityScaling3} we find that, to leading order in $1/t_f$,
\begin{gather}
	\begin{aligned}\label{eq:ClassB_ED}
	&\mathcal{E}(t_f,\gamma,\kappa,\dmu_i,T_i)\\
	&=(t_0/t_f)^\zetaB\mathcal{D}(\gamma,\bar{\kappa}_\infty,\bar{v}_\mu=t_0^{-\beta}\dmu_i,\bar{v}_T=t_0^{-\alpha}T_i)
	\end{aligned}
\end{gather}
where
\begin{equation}
	\label{eq:ClassB_ScalingPower}
	  \zetaB\equiv\frac{\alpha}{z(1+s\alpha)}=\frac{\nu z\beta}{z(1+s\nu z\beta)}.
\end{equation}
The second equivalence holds since $\alpha=\beta\nuz$ for ramps in this class.

\subsubsection{Ramps in class C}
Class C on the one hand covers $\mu$-only ramps for which $T_i=0$, and on the other hand simultaneous $T$-$\mu$ ramps with $\alpha/\beta>\nuz$. Here we use Scheme 2, and set $r$ and $p$ as in \eqref{eq:Scheme2_rp}. From \eqref{eq:Scheme2_speeds} it is found that $\lim_{f,q\rightarrow\infty}\bar{v}_T=0$. Again from Eq.~\eqref{eq:ExcitationDensityScaling3} we find to leading order in $1/t_f$ that
\begin{align}
		&\mathcal{E}(t_f,\gamma,\kappa,\dmu_i,T_i)\nonumber\\
		&=(t_0/t_f)^\zetaC\mathcal{D}(\gamma,\bar{\kappa}_\infty,\bar{v}_\mu=t_0^{-\beta}\dmu_i,\bar{v}_T=0)\label{eq:ClassC_ED}
\end{align}
where
\begin{equation}
	\label{eq:ClassC_ScalingPower}
	\zetaC\equiv\frac{\nu z\beta}{z(1+s\nu z\beta)}.
\end{equation}
The excitation density $\mathcal{D}$ on the right of Eq.~\eqref{eq:ClassC_ED} is that of an infinite-time ramp for which the velocity associated with the variation in $T$ is zero. Therefore $\mathcal{D}$ corresponds to a $\mu$-only ramp. This is consistent with the observation that the dependence on $T_i$ has dropped out in \eqref{eq:ClassC_ED}. In the slow-ramp limit all ramps in class C are therefore equivalent to $\mu$-only ramps.

\subsubsection{Scaling with ramp velocities}
\label{sec:ScalingWithRampVelocities}
The results for the scaling of $\mcE$ can also be rephrased in terms of the ramp velocities $v_\mu$ and $v_T$. Consider a class B ramp with a particular starting point $(\dmu_i,T_i)$, for which the scaling behavior is given in Eq.~\eqref{eq:ClassB_ED}. We can eliminate $t_0$ in favor of $\bar{v}_\mu=t_0^{-\beta}\dmu_i$ or $\bar{v}_T=t_0^{-\alpha}T_i$. This leads to the two equivalent forms
\begin{align}
	\mcE&=\left(v_T/\bar{v}_T\right)^{\zetaB/\alpha}\mathcal{D}(\gamma,\bar{\kappa}_\infty,\bar{v}_\mu,\bar{v}_T),\label{eq:EitovT}\\
	\mcE&=\left(v_\mu/\bar{v}_\mu\right)^{\zetaB/\beta}\mathcal{D}(\gamma,\bar{\kappa}_\infty,\bar{v}_\mu,\bar{v}_T),\label{eq:Eitovmu}
\end{align}
that capture the scaling of the excitation density with the individual ramp velocities. It should be kept in mind that the velocities $v_\mu$ and $v_T$, and similarly $\bar{v}_\mu$ and $\bar{v}_T$, cannot be varied independently. In fact, these are required to satisfy $v_T^\beta/v_\mu^\alpha=\bar{v}_T^\beta/\bar{v}_\mu^\alpha=T_i^\beta/\dmu_i^\alpha$. This constraint encodes the precise curve that the ramp path follows in the $\dmu$-$T$ plane, which is defined by $T^\beta/\dmu^\alpha=T_i^\beta/\dmu_i^\alpha$. Shifting the starting point of the ramp along this curve would not affect the value of the ratio $T_i^\beta/\dmu_i^\alpha$, and in this sense $\mcE$ in Eqs.~\eqref{eq:EitovT} and \eqref{eq:Eitovmu} depends on the curve that the ramp path follows, but not on the ramp's starting point. This is to be expected, since the system initially evolves adiabatically, and it is only the approach to the critical point at late times that impacts $\mcE$ at leading order. The same arguments can be applied to ramps in classes A and C, leading to the scaling laws
\begin{align}
	\text{Class A:}\  \mcE&=\left(v_T/\bar{v}_T\right)^{\zetaA/\alpha}\mathcal{D}(\gamma,\bar{\kappa}_\infty,0,\bar{v}_T),\\
	\text{Class C:}\  \mcE&=\left(v_\mu/\bar{v}_\mu\right)^{\zetaC/\beta}\mathcal{D}(\gamma,\bar{\kappa}_\infty,\bar{v}_\mu,0).
\end{align}

\begin{figure*}[t]
    \subfloat[\label{fig:ScalingOhmic}]{%
  \includegraphics[width=0.48\textwidth]{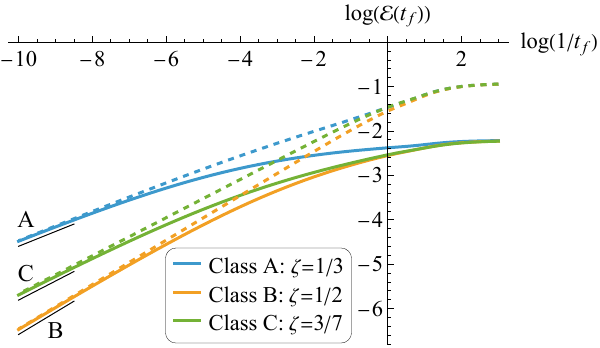}%
}\hfill
    \subfloat[\label{fig:EDvsvT0}]{%
  \includegraphics[width=0.48\textwidth]{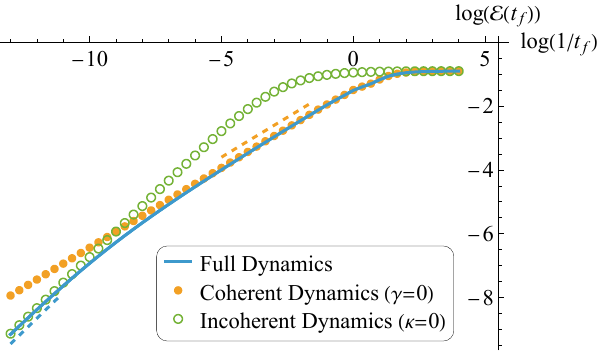}%
}
	\caption{\label{fig:EDvsv} (a) Numerical results for the logarithm of the excitation density $\mcE=\mcE(t_f)$ at the end of a ramp toward the critical point, shown as a function of $\log(1/t_f)$. Results were obtained using Eq.~\eqref{eq:ExcitationDensity} and  solving the dynamical equations in \eqref{eq:FullDynamics} numerically for the Kitaev chain. Results for the three ramp classes A, B, and C appear in different colours. For class A we chose $(\alpha,\beta)=(1/2,1)$, for class B $(\alpha,\beta)=(1,1)$, and for class C $(\alpha,\beta)=(1,3/4)$. Solid lines indicate results obtained using the exact expressions for $\lambda(\dmu,k)$ and $\beta(\dmu,k)$ in Eq.~\eqref{eq:KitaevLambdaBeta}. Dashed lines are results obtained using the local approximations in Eq.~\eqref{eq:KitaevLocalApproximations}. Solid black lines have slopes corresponding to the predicted value of $\zeta$ for each of the three ramps. 
    Parameter values are $J=\Delta_p=1$, $s=\kappa=\delta=1$, $T_i=\dmu_i=0.28$, $\gamma=0.05$. (b) As in (a), but for a $\mu$-only ramp at $T=0$, starting from $\mu_i=-5$ and ending at $\mu_c=-1$. The bath is sub-ohmic with spectral parameter $s=1/4$. Here the local approximations in Eq.~\eqref{eq:KitaevLocalApproximations} were used, and we set $\alpha=\beta=1$. Also shown is data obtained by restricting to purely coherent ($\gamma=0$) and incoherent ($\kappa=0$) dynamics. The orange and blue dashed lines have slopes corresponding to the scaling exponents $\zetaC=1/2$ and $\zetaCoh=4/5$, respectively. Remaining parameters are $J=\Delta_p=1$, $\gamma=0.02$, and $\delta=J^{1-s}$.}
\end{figure*}

\subsection{Discussion and Numerical Results}
\label{sec:DiscussionResults}

We check the predictions for the scaling behavior of the excitation density derived in the previous section against numerical data for the Kitaev chain defined in Sec.~\ref{sec:KitaevDefinition}.  Figure \ref{fig:ScalingOhmic} shows exact numerical results for the final excitation density $\mathcal{E}(t_f)$ as a function of the inverse ramp time $1/t_f$. These results are obtained by solving the dynamical equations in \eqref{eq:FullDynamics} numerically and inserting the results into Eq.~\eqref{eq:ExcitationDensity}. The slopes of the solid black lines in Fig.~\ref{fig:ScalingOhmic} correspond to the predicted value of $\zeta$ for each of the three ramps shown. The straight-line behavior of $\log(\mathcal{E}(t_f))$ versus $\log(1/t_f)$ at large $t_f$ confirms power-law scaling, and we observe very good agreement with the predicted scaling powers. Solid curves were obtained using the exact expressions for the dispersion relation and Bogoliubov angle, while the dashed curves are based on the local approximations in Eq.~\eqref{eq:KitaevLocalApproximations}. As expected, the two approaches agree for sufficiently slow ramps.

As a second example we consider a ramp protocol that aligns closely with the traditional Kibble-Zurek setup. We set $T_i=0$, and ramp $\mu$ toward its critical value $\mu_c$ as per Eq.~\eqref{eq:RampDefinition}. This amounts to a pure parameter ramp at zero temperature, with the system initialized in its ground state. In the absence of the bath coupling, i.e., with $\gamma=0$, the system evolves unitarily, and the final excitation density exhibits Kibble-Zurek scaling in $1/t_f$ with a power of \cite{Barankov08,Sen08,DeGrandi10}
\begin{equation}
	\label{eq:CoherentScalingPower2}
	\zetaCoh=\frac{\nu\beta}{1+\nuz\beta}.
\end{equation}
A derivation of this result using our approach appears in Appendix \ref{app:CoherentScaling}. In the presence of a bath, the predicted scaling exponent is that of a class C ramp as given in Eq.~\eqref{eq:ClassC_ScalingPower}. For ohmic baths ($s=1$) these two exponents are equal. However, this does not imply that the excitation dynamics is unaffected by the bath, but only that it is the value of $\mathcal{E}(t_f)$, rather than its scaling with $1/t_f$, that gets modified. In contrast, for sub-ohmic baths ($s<1$) the exponents in Eqs.~\eqref{eq:CoherentScalingPower2} and \eqref{eq:ClassC_ScalingPower} differ, and the bath coupling actually affects the scaling of $\mathcal{E}(t_f)$. One might wonder how this modified scaling comes about, since for any fixed ramp time $t_f$ the impact of the bath on the value of $\mcE(t_f)$ can be made arbitrarily small by choosing a sufficiently weak bath coupling $\gamma$. This observation suggests that $\mcE(t_f)$ should exhibit a cross-over in its scaling behavior as $1/t_f$ is decreased. Specifically, for $1/t_f$ either well above or well below this crossover, it is, respectively, the coherent dynamics or the bath coupling that will dominate, leading to scaling according to either $\zetaCoh$ or $\zetaC$. This is illustrated in Fig.~\ref{fig:EDvsvT0}, where we show numerical data for $\log(\mcE(t_f))$ as a function of $\log(1/t_f)$. Also included is data for the purely coherent ($\gamma=0$) and incoherent ($\kappa=0$) cases. The predicted cross-over behavior between coherent and bath-dominated dynamics is clearly visible. The orange and blue dashed lines have slopes corresponding to the scaling powers $\zetaCoh$ and $\zetaC$, respectively. Overall, the numerical checks convincingly confirm the asymptotic validity of the scaling laws derived in this article.

When simultaneously varying $T$ and $\mu$, the dynamics of the excitation density is affected by coherent as well as incoherent processes in the master equation \eqref{eq:LindbladMasterEquation}. For an in-depth understanding of the shapes of the curves in Fig.~\ref{fig:EDvsv} beyond their asymptotics, it is instructive to disentangle the complicated interplay between the generation of excitations due to varying the system Hamiltonian, a partial suppression of these excitations by the unitary dynamics, as well as dissipation or heating due to the coupling with the thermal environment. Such an analysis is presented in Appendix \ref{app:DynamicalProcesses}.

\begin{figure*}
	\centering
	  \includegraphics[width=1\linewidth]{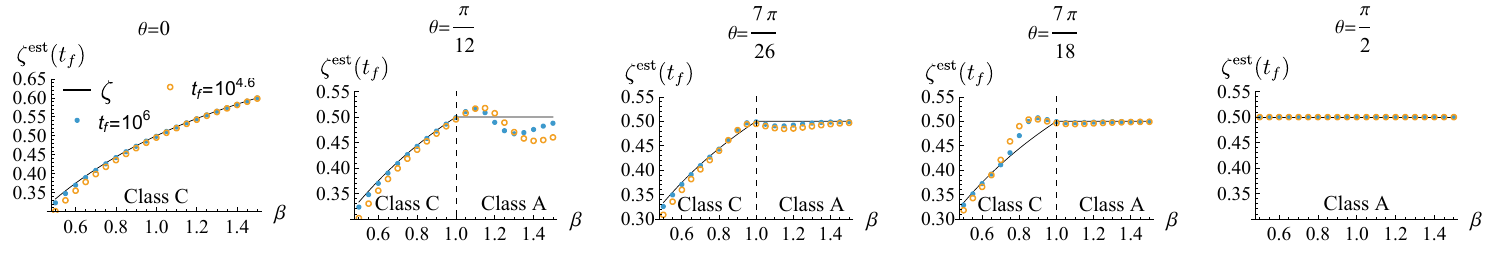}
	\caption{A comparison of the scaling exponent $\zeta$ calculated in Sec.~\ref{sec:ScalingLaws} (solid black line) with the exponent $\zeta^{\rm est}(t_f)$ extracted numerically using Eq.~\eqref{eq:zetaest} from data for the Kitaev chain. Results for $t_f=10^6$ (blue dots) and $t_f=10^{4.6}$ (orange circles) are shown for a range of $\beta$ values, with $\alpha=1$ fixed. The ramp starts at $(\dmu_i,T_i)=(-4\cos(\theta),4\sin(\theta))$, and results for five values of $\theta$ are shown. Other parameters are $J=\Delta_p=1$, $s=\kappa=\delta=1$ and $\gamma=1/15$.}
	\label{fig:zetavsbeta}
\end{figure*}

\subsection{Probing critical exponents using dynamical data}
\label{sec:extraction}
The central notion of the original Kibble-Zurek framework, and of our finite-temperature extension thereof, is that properties of an equilibrium phase transition can be imprinted on a system's nonequilibrium dynamics. This is reflected by the scaling laws for the excitation density $\mcE(t_f)$ derived in Sec.~\ref{sec:ScalingLaws}, which carry a signature of the phase transition via the critical exponents $\nu$ and $z$. Knowledge of these exponents therefore enables predictions of this scaling behavior with respect to the ramp time or a chosen ramp velocity. The reverse is also true: an experimental realization of the ramping protocols can function as a probe of the universality class of the equilibrium phase transition by allowing the critical exponents to be extracted from dynamical data. In this section we highlight, within the context of our open-system model, some general considerations that may enter in the choice or design of such a protocol. Of course, its practical limitations and ultimate feasibility will depend on the details of the system under consideration and on the level of control that can be achieved over the system and environmental parameters.
\begin{figure}[h]
	\centering
	\includegraphics[width=1\linewidth]{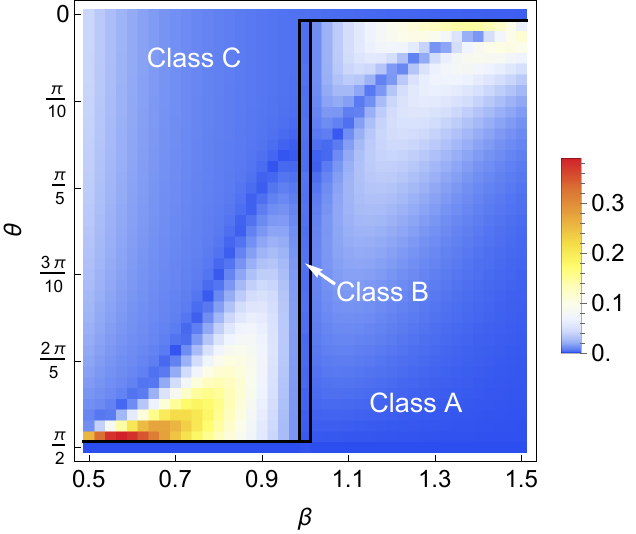}
	\caption{The relative difference $(\zeta^{\rm est}(t_f)-\zeta)/\zeta$ between the scaling exponent $\zeta$ derived in Sec.~\ref{sec:ScalingLaws} and the estimate $\zeta^{\rm est}(t_f)$ calculated numerically for the Kitaev chain using Eq.~\eqref{eq:zetaest} for $41$ values of $\beta$ and of $\theta$. Here $\nu=z=1$ and $t_f=\exp(10.7)$. Regions bounded by black lines correspond to data points for different ramp classes, with the central rectangular region, where $\beta=1$, corresponding to class B ramps. Note that the size of the latter region, and that of class A ramps with $\beta\leq1$ and class C ramps with $\beta\geq1$ appear exaggerated in the figure due to the discrete set of $\theta$ and $\beta$ values used in the calculation. See the caption of Fig.~\ref{fig:zetavsbeta} for parameter settings and the definition of $\theta$.}
	\label{fig:zetaerror}
\end{figure}
Regardless of the precise setting, it is generally favorable to employ ramping protocols for which the predicted asymptotic scaling emerges rapidly with decreasing ramp velocities, i.e., at shorter rather than longer ramp times, so as to minimize the time over which system and environment need to be controlled. This favors ramps for which $\mcE(t_f)$ converges quickly to its asymptotic form \eqref{eq:ExcitationDensityScaling2} as $t_f$ increases. We expect that the convergence rate will be affected by the ramp class as well as the choice of the starting point $(\dmu_i,T_i)$. Determining the convergence rate analytically would require deriving scaling laws for the next-to-leading-order terms in the asymptotic expansion of $\mcE(t_f)$ in $1/t_f$, a problem that lies outside the scope of the present article. Instead, we proceed numerically and solve the full set of dynamical equations \eqref{eq:FullDynamics} for the Kitaev chain over a range of ramp parameters. From these numerical data, we extract finite-$t_f$ estimates for the scaling exponent $\zeta$ using
\begin{equation}\label{eq:zetaest}
    \zeta^{\rm est}(t_f)=-\frac{{\mathrm d}\log\mcE(t_f)}{{\mathrm d}\log(t_f)}.
\end{equation}
The degree to which $\zeta^{\rm est}(t_f)$ matches the \emph{asymptotic} exponent $\zeta$ encoding $z$ and $\nu$, as derived in Sec.~\ref{sec:ScalingLaws}, provides an indication of the rate of convergence to the asymptotic scaling law \eqref{eq:ExcitationDensityScaling2}. Figure \ref{fig:zetavsbeta} shows this estimate $\zeta^{\rm est}(t_f)$, for $t_f=10^{4.6}$ and $t_f=10^6$, over a range of $\beta$ values, with $\alpha=1$ fixed. The solid lines in the plots indicate the asymptotic $\zeta$ exponents calculated in Sec.~\ref{sec:ScalingLaws}. The subplots correspond to ramps with different starting points, parametrized by the angle $\theta$ as $(\dmu_i,T_i)=(-4\cos(\theta),4\sin(\theta))$. As expected, exponents extracted at longer ramp times match the asymptotic values more closely. The deviation of $\zeta^{\rm est}(t_f)$ from $\zeta$ depends strongly on both the ramp class and the starting point of the ramp. This is also illustrated in Fig.~\ref{fig:zetaerror}, which shows the relative difference between $\zeta^{\rm est}(t_f)$ and the predicted $\zeta$ for a range of $\beta$ and $\theta$ values. For pure parameter and temperature ramps (so for $\theta=0$ and $\theta=\pi/2$) the predicted scaling behavior emerges comparatively quickly over the full range of $\beta$ values, in contrast to what is seen for ramps starting from within the $\dmu$-$T$ plane. We vary $\beta$ with $\alpha=1$ fixed, but have confirmed that varying $\alpha$ yields qualitatively similar results.

The results displayed in Fig.~\ref{fig:zetaerror} provide guidance for choosing efficient ramp protocols for extracting accurate equilibrium critical exponents from nonequilibrium data. Three regions in the plot catch the eye, where blue colors indicate rapid convergence of the excitation density to the predicted scaling behavior:
\begin{enumerate}
\item Ramps with $\theta=\pi/2$, independent of the ramp exponent $\beta$. These are $T$-only ramps at constant $\mu=\mu_c$, discussed in detail in Ref.~\cite{KingKrielKastner23}. While being simple and efficient, these ramps have the drawback of giving access only to the exponent $z$, but not $\nu$.
\item Ramps with $\theta=0$, independent of the ramp exponent $\beta$. These are $\mu$-only ramps at $T=0$. If the scaling exponent can be extracted for two different values of $\beta$, then this would be sufficient to determine $z$ and $\nu$ when the bath spectral parameter $s$ is known. Comparable agreement between $\zeta^{\rm est}(t_f)$ and $\zeta$ is found for class A ramps starting close to the $T$ axis ($\theta\approx\pi/2$) and class C ramps starting close to the $\dmu$ axis ($\theta\approx0$). 
\item Class B ramps, which, for the parameter settings used in Fig.~\ref{fig:zetaerror}, correspond to $\beta=1$. Regardless of the starting point set by $\theta$, and even at ramp times for which the agreement found for classes A and C is much poorer, class B ramps yield high-accuracy estimates of the scaling exponent $\zeta$. Implementing class B ramps, however, requires knowledge of the critical exponents, since the ratio of powers $\alpha/\beta$ must be tuned to $\nu z$. If $\nu z$ is unknown, this may be a problem. If, however, the goal is to \emph{verify} the predicted scaling for simultaneous parameter and temperature ramps toward a critical point with known critical exponents, then class B ramps offer some unique advantages.
\end{enumerate}
In the above discussion we assumed full control of the parameters $\mu$ and $T$. Experimental constraints may of course impose limitations, and other types of protocols may be optimal when taking these constraints into account.

\section{Summary and Conclusions}
\label{sec:SummaryConclusions}

Temperature ramps offer exciting opportunities for probing, via the Kibble-Zurek framework, quantum critical properties in open systems at nonzero temperatures \cite{KingKrielKastner23}, but suffer from the drawback that only the dynamical critical exponent $z$ can be probed, and not $\nu$. This shortcoming was overcome in the present article by generalizing to nonlinear, simultaneous ramps of the bath temperature and a Hamiltonian parameter. Working with free fermionic chains coupled to thermalizing baths, the main technical achievement of our article is a leading-order (in the ramp speed) asymptotic characterization of the chain's excitation density at the end of a ramp toward the quantum critical point. This excitation density $\mcE(t_f)$ exhibits power-law scaling in the ramp time $t_f$, with a scaling exponent $\zeta$ that is a function of equilibrium critical exponents as well as the bath spectral exponent $s$.

In general the value of the scaling exponent $\zeta$ depends smoothly on the exponents $\alpha$ and $\beta$ that characterize the ramping path in the $\dmu$-$T$ plane. Only when $\alpha/\beta=\nu z$ is there a nonanalytic transition between scaling laws, from what we called class A scaling exponents \eqref{eq:ClassA_ScalingPower} on the one side, to class C scaling exponents \eqref{eq:ClassC_ScalingPower} on the other. In physical terms, class A ramping paths asymptotically approach the quantum critical point through the quantum critical region, whereas class C paths approach from the semi-classical side, as illustrated in Fig.~\ref{fig:ramp-paths}. We found that class A scaling exponents $\zeta_A$, identical to those of $T$-only ramps, depend on the critical exponent $z$, but not on $\nu$. Class C scaling exponents $\zeta_C$, however, depend on both $z$ and $\nu$.

A potential application of the class C scaling laws is to extract the critical exponents $z$ and $\nu$, and hence characterize the universality class of the equilibrium quantum phase transition from nonequilibrium data. Key to this task is an understanding of the size of the corrections to the asymptotic scaling laws: small corrections imply that the desired scaling laws are manifest already at higher ramp velocities, i.e., shorter ramp times, which is likely to be beneficial in experiments.

A numerical investigation of this issue, summarized in Fig.~\ref{fig:zetaerror}, identifies qualitative trends that can inform the design of an experimental protocol for probing critical properties in nonequilibrium conditions. Class B ramps, where path exponents $\alpha$ and $\beta$ are chosen such that $\alpha/\beta=\nu z$, are found to be particularly advantageous, as their leading asymptotic scaling behavior already emerges at rather high ramp velocities and largely independent of the starting point of the ramp.

Several technical aspects and further generalizations had to remain unaddressed in this article, but may stimulate future research. This includes an analytic investigation of the convergence rate of the excitation density to its asymptotic scaling form; a derivation of scaling laws in the case of open-system dynamics induced by a super-ohmic bath ($s>1$); and the search for suitable nonlocal observables that reveal information about the topological features of the quantum phase transition of the Kitaev chain under nonequilibrium conditions and at nonzero temperatures.

The main open question, and certainly an ambitious one, is whether the conclusions of this article carry over to interacting fermionic or bosonic models, or spin models. We do not have preliminary results or other insights to offer in this regard, but can only express our expectation: at least in the traditional context of closed quantum systems, a key feature of Kibble-Zurek physics is precisely its model independence. What matters is not the details, but only the universal properties close to the critical point of the model. Whether this model independence persists in the case of simultaneous $T$-$\mu$ ramps in open systems remains to be investigated.

\appendix
\renewcommand\thefigure{\thesection\arabic{figure}}    
\counterwithin{figure}{section}

\section{Other rescaling schemes}
\label{app:RescalingSchemes}
We define the two additional schemes that fulfill the criteria set out in Sec.~\ref{sec:StrategyOverview}. The second of these (scheme 4) will be used in Appendix \ref{app:CoherentScaling} to derive scaling results for ramps under purely unitary dynamics. Both these schemes produce a $\bar{\kappa}=\kappa$ parameter that remains fixed, i.e., independent of the rescaling. They differ in terms of the ramp velocities associated with $\bmcE$. For scheme 3, the velocity $\bar{v}_T$ is fixed, while $\bar{v}_\mu$ is fixed for scheme 4. The details are as follows:
\begin{itemize}[leftmargin=*]
	\item \textbf{Scheme 3:} We set
	\begin{equation}
		\label{eq:Scheme3_rp}
		r=p/s=\frac{\alpha}{1+\alpha},\qquad q=f^{1-r}.
	\end{equation}
    Referring back to Eqs.~\eqref{eq:RescaledParameters} and \eqref{eq:RampsSpeeds2}, the parameters of $\bmcE$ are
	\begin{align}
	&(\bar{t}_f,\bar{\gamma},\bar{\kappa},\widebar{\dmu}_i,\widebar{T}_i)\nonumber\\
	&=(qt_0,q^{\alpha(1-s)}\gamma,\kappa,q^{\alpha/(\nuz)}\dmu_i,q^{\alpha} T_i),\label{eq:Scheme3_parameters}
	\end{align}
	with ramp velocities
	\begin{equation}
		\label{eq:Scheme3_speeds}
	\bar{v}_T=t_0^{-\alpha}T_i,\qquad \bar{v}_\mu=q^{\alpha/(\nuz)-\beta}t_0^{-\beta}\dmu_i.
	\end{equation}
	\item \textbf{Scheme 4:} We set
	\begin{equation}
		\label{eq:Scheme4_rp}
	r=p/s=\frac{\nuz\beta}{1+\nuz\beta},\qquad q=f^{1-r}.
	\end{equation}
	The parameters of $\bmcE$ are
	\begin{align}
		&(\bar{t}_f,\bar{\gamma},\bar{\kappa},\widebar{\dmu}_i,\widebar{T}_i)\nonumber\\
		&=(qt_0,q^{\nuz\beta(1-s)}\gamma,\kappa,q^\beta\dmu_i,q^{\nuz\beta} T_i),\label{eq:Scheme4_parameters}
	\end{align}
	with ramps velocities
	\begin{equation}
		\label{eq:Scheme4_speeds}
		\bar{v}_T=q^{\nuz\beta-\alpha}t_0^{-\alpha}T_i,\qquad \bar{v}_\mu=t_0^{-\beta}\dmu_i.
	\end{equation}
\end{itemize}

\section{Dynamical processes}
\label{app:DynamicalProcesses}

\begin{figure}[t!]
  \includegraphics[width=1\linewidth]{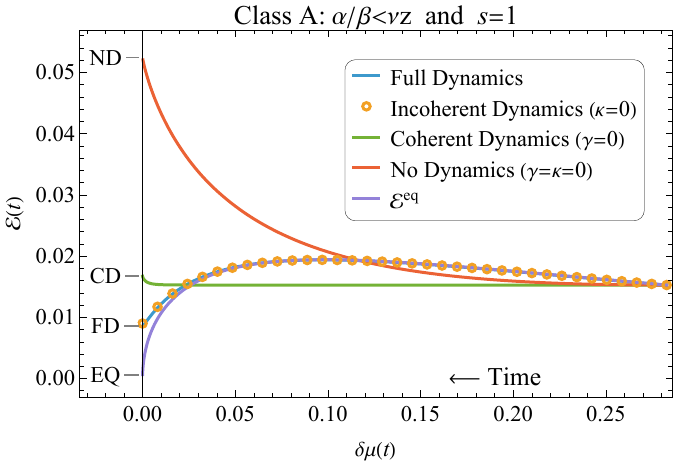}\\
  \includegraphics[width=1\linewidth]{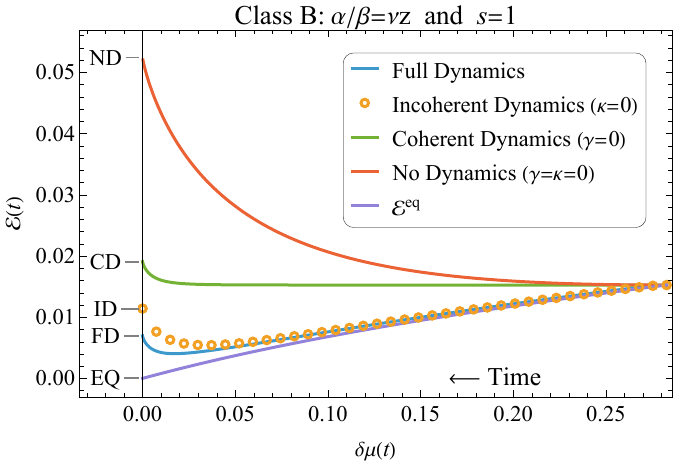}\\
  \includegraphics[width=1\linewidth]{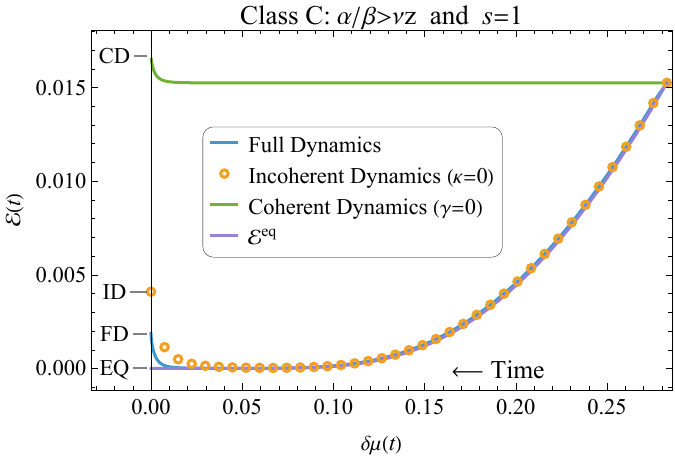}
    \caption{Numerical results for the excitation density $\mcE(t)$ as a function of $\dmu(t)$ for a class A $(\alpha=1/2,\beta=1)$, class B \mbox{$(\alpha=1,\beta=1)$}, and class C ($\alpha=2$, $\beta=1$) ramp. The bath is ohmic with $s=1$. Curves corresponding to various choices of $\kappa$ and $\gamma$ are shown in each plot, together with the instantaneous thermal equilibrium value of the excitation density. All data are for the Kitaev chain defined in Sec.~\ref{sec:KitaevDefinition} with parameters $J=\Delta_p=1$, $\kappa=\delta=1$, $T_i=\dmu_i=0.28$, $\gamma=0.05$ and $t_f=\exp(5)$ for class A, and $t_f=\exp(7)$ for class B and C. All ramps start at $\mu_i=-1.28$ and end at $\mu_f=\mu_c=-1$.}
	\label{fig:ClassABCvsT}
\end{figure}

Here we revisit the discussion in Sec~\ref{sec:CorrelationDynamics} of the various processes contained in the dynamical equations \eqref{eq:FullDynamics}, now in the context of ramps toward the critical point in the Kitaev chain of Sec.~\ref{sec:KitaevDefinition}. We will consider the role of these processes in determining both the value and the scaling behavior of the excitation density for the various ramp classes. It is important to note that the bath coupling is essential for producing the scaling behavior found in Sec.~\ref{sec:ScalingLaws}, as it ensures that the final excitation density $\mathcal{E}(t_f)$ tends to zero in the adiabatic limit $t_f\rightarrow\infty$. In the absence of the bath, i.e., for purely coherent dynamics, the adiabatic limit produces a \emph{time-independent} excitation density, and therefore $\mathcal{E}(t_f)$ would tend to $\mathcal{E}(0)$. In this case $\mathcal{E}(t_f)$ cannot exhibit power-law scaling at leading order in $1/t_f$, except when the system is initialized in its ground state, for which $\mathcal{E}(0)=0$. This is the scenario considered in the traditional Kibble-Zurek protocol. For positive $\mathcal{E}(0)$, the appropriate quantity to consider in the unitary case is the \emph{non-adiabatic correction} $\mathcal{E}(t_f)-\mathcal{E}(0)$, which exhibits the standard Kibble-Zurek-type scaling, as demonstrated in Appendix \ref{app:CoherentScaling}. 

For ramps from each of the three classes defined in Sec.~\ref{sec:Ramps}, we calculated the total excitation density $\mcE(t)$ as a function of time by solving the dynamical equations \eqref{eq:FullDynamics} numerically and inserting the outcome into Eq.~\eqref{eq:ExcitationDensity}. Setting $\kappa$ and/or $\gamma$ to zero allows us to isolate particular processes. The results are shown in Fig.~\ref{fig:ClassABCvsT}. Note that $\dmu(t)$ appears on the horizontal axis, and so time can be regarded as increasing from right to left. The data was produced for the case of an ohmic bath, where $s=1$. We return to the sub-ohmic case at the end. 

\subsection{Class B}
Data for an example of a class B ramp is shown in Fig.~\ref{fig:ClassABCvsT} (center). The top curve corresponds to setting both $\kappa$ and $\gamma$ to zero. This is equivalent to setting the right-hand side of the master equation \eqref{eq:LindbladMasterEquation} to zero, resulting in the system's density matrix being constant in time. The change in $\mathcal{E}(t)$ is now solely due to the explicit time dependence of the Hamiltonian and the resulting change in its instantaneous eigenstates. To be clear, $\rho(t)=\rho(0)$ remains unchanged, but the $\eta$-fermions that define $P_k=\ave{\eta_k^\dag \eta_k^\pdag}$ are defined from the original $c$-fermions in the Hamiltonian through the time-dependent unitary transformation in Eq.~\eqref{eq:bogo}. As a result, the mode occupations $P_k$ increase with time. In Eq.~\eqref{eq:FullDynamics} this is brought about by the ${\sim}{\rm d}\beta_k/{\rm d}t$ terms. While this situation does not represent a physical scenario, it serves as a reference case for understanding the impact of unitary dynamics and bath coupling.

Setting $\kappa=1$ but keeping $\gamma=0$ results in purely unitary dynamics, albeit starting from a thermal state. Away from the critical point where the excitation gap $\Delta\sim|\dmu|$ is large, the system evolves adiabatically, and hence the excitation density remains approximately constant at its initial value. This is illustrated by the green curve in Fig.~\ref{fig:ClassABCvsT} (center). Close to the critical point, we see a rapid increase in the excitation probability, which indicates a breakdown of adiabaticity. 

Setting $\kappa=0$ and $\gamma>0$ couples the system to the bath, while neglecting the unitary $-i[H,\rho]$ term in the master equation. The notion of adiabaticity is now a different one and corresponds to the system's ability to remain in thermal equilibrium with the bath. Away from the critical point where the mode relaxation rates $R(\lambda,T)>2\pi\gamma\,\text{max}\{\lambda^s,(2T)^s\}$ are large, the excitation density in orange indeed tracks its thermal equilibrium value shown in purple. However, close to the critical point the slow relaxation rates of the low-energy modes combined with the excitations generated by the Hamiltonian driving result in the system dropping out of equilibrium with the bath.

Finally, we consider the full dynamics with $\kappa=1$ and $\gamma>0$ shown in blue. Early in the ramp, the excitation density closely follows that of the incoherent case, suggesting that the dynamics is dominated by the bath coupling. Closer to the critical point the unitary dynamics starts to counteract the generation of excitations by the Hamiltonian driving, leading to a final excitation density that is lower than that of the purely incoherent case. From this behavior, it is evident that coherent as well as incoherent processes contribute nontrivially to the final excitation density.

\subsection{Class A}
As shown in Sec.~\ref{sec:RampsA}, class A ramps yield the same excitation density as temperature-only ramps to leading order in $1/t_f$. As a result, the slow-ramp dynamics of the excitation density is dominated by the bath coupling and can be described by the rate equation for $P_k$ that follows from setting ${\rm d}\beta_k/{\rm d}t$ to zero in \eqref{eq:FullDynamics}. This is supported by the numerical results shown in Fig.~\ref{fig:ClassABCvsT} (top). Here it is seen that the evolution of $\mathcal{E}$ under the full dynamics (shown in blue) is essentially identical to that of the purely incoherent dynamics (shown in orange) that result from setting $\kappa=0$.

\subsection{Class C}
For class C ramps both the coherent and incoherent dynamics play a role in determining the final excitation density, as was also the case for class B ramps. This is illustrated Fig.~\ref{fig:ClassABCvsT} (bottom), where the value of $\mcE{(t_f)}$ that follows from the full dynamics is seen to differ significantly from that of the purely incoherent case.

\subsection{Sub-ohmic baths}
For sub-ohmic baths ($s<1$), the dynamics for ramp classes B and C is modified and the relaxation rate $R\sim\lambda^s$ in Eq.~\eqref{eq:RDefinition} is amplified for the low-lying modes. As a result, the incoherent dynamics becomes dominant for all three classes of ramps. This is supported by the fact that the $\mathcal{D}$ excitation densities in Eqs.~\eqref{eq:ClassA_ED}, \eqref{eq:ClassB_ED}, and \eqref{eq:ClassC_ED} are characterized by a rescaled $\bar{\kappa}=0$ and are therefore independent of $\kappa$ at leading order in $1/t_f$. Put differently, setting $\kappa=0$ and thereby dropping the unitary $-i\left[H,\rho\right]$ term from the master equation \eqref{eq:LindbladMasterEquation} has no effect on $\mcE(t_f)$ to leading order.

\section{Scaling results for unitary dynamics}
\label{app:CoherentScaling}
The approach used in this paper also yields results for ramps where the system undergoes purely unitary dynamics, i.e., with the bath coupling $\gamma$ set to zero. Despite the absence of a bath in the dynamics, we will initialize the system in a thermal equilibrium state at temperature $T_i$. This allows for a more direct comparison with other results. When $T_i=0$ this amounts to starting the system off in its ground state, which should reproduce known results for the standard Kibble-Zurek protocol. As emphasized in App.~\ref{app:DynamicalProcesses}, in the adiabatic limit $t_f\rightarrow\infty$ the final excitation density $\mcE(t_f)$ does not vanish (as it does when $\gamma>0$), but instead tends to its initial value $\mcE(0)$. When the latter is positive, non-trivial scaling behavior can only appear in the nonadiabatic correction $\mcE(t_f)-\mcE(0)$. To isolate this correction, we return to Eq.~\eqref{eq:FullDynamics}, now with $\gamma$ set to zero and the $k$-subscript suppressed:
\begin{subequations}
	\label{eq:FullDynamicsUnitary}
	\begin{align}
		\deriv{P}{t}&=2\deriv{\beta}{t} {\rm Im}(C),\\
		\deriv{C}{t}&=2i\kappa\lambda C-2i\deriv{\beta}{t}(P-1/2).
	\end{align}
\end{subequations}
Suppose $P$ and $C$ satisfy these equations with initial conditions $P=P(0)$ and $C=0$. From this solution, we construct a second solution \mbox{$\tilde{P}=uP+(1-u)/2$}, $\tilde{C}=u C$ for which the initial conditions are $\tilde{P}(0)=uP(0)+(1-u)/2$ and $\tilde{C}(0)=0$. This holds for any choice of $u$. If we take $u=1/(1-2P(0))$, then $\tilde{P}(0)=0$. This shows how the excitation probability $P(t_f)$ following from a nonzero initial condition $P(0)>0$ can be related to the probability $\tilde{P}(t_f)$ associated with a zero initial condition $\tilde{P}(0)=0$. In other words, we can relate the excitation probabilities for a ramp with $T_i>0$ to those of a ramp with $T_i=0$. Writing $P(t,T_i)$ for $P$, this amounts to
\begin{equation}\label{eq:PRewrite}
	P(t_f,T_i)=P(t_f,0)(1-2P(0,T_i))+P(0,T_i).
\end{equation}
Here $P(0,T_i)=P^{\rm th}(\lambda(\dmu_i,\epsilon)/T_i)$ is the thermal equilibrium (Fermi-Dirac) excitation probability. Inserting Eq.\ \eqref{eq:PRewrite} into the integral expression for $\mathcal{E}(t_f)$ in \eqref{eq:ExcitationDensityEnergyIntegralLocal} gives
\begin{align}
&\mathcal{E}(t_f)-\mathcal{E}(0)\!=\!\!\int_0^{\infty}\!\!{\rm d}\epsilon[\tilde{\rho}(\epsilon)(1-2P^{\rm th}(\lambda(\dmu_i,\epsilon)/T_i))\nonumber\\
&\qquad\times P(\epsilon,t_f,\gamma=0,\kappa,\dmu_i,T_i=0)].
\end{align}
The excitation probability \mbox{$P(\epsilon,\ldots,T_i=0)$} in the integrand above is invariant under rescaling in Eq.~\eqref{eq:ExcitationDensityScaling1}. We choose $a$ and $b$ as in Eq.~\eqref{eq:abfDefinition}, and $p$ and $r$ according to Scheme 4 as per Eq.~\eqref{eq:Scheme4_parameters}. Applying this rescaling to the excitation probability and then rescaling the integration variable $\epsilon\rightarrow\epsilon/a$ leads to
\begin{equation}
	\label{eq:CorrectionScaling}
	\mathcal{E}(t_f)-\mathcal{E}(0)=(t_0/t_f)^\zeta\bmcE
\end{equation}
with 
\begin{equation}
	\label{eq:CoherentScalingPower}
	\zeta=\zetaCoh=\frac{\nu\beta}{1+\nuz\beta}
\end{equation} and
\begin{align}
	&\bmcE=\int_0^\infty\!{\rm d}\epsilon[\tilde{\rho}(\epsilon)(1-2P^{\rm th}(\Lambda(\dmu_i,(t_0/t_f)^{p/s}\epsilon)/T_i))\nonumber\\
	&\qquad\times P(\epsilon,qt_0,\gamma=0,\kappa,q^\beta\dmu_i,T_i=0)]\label{eq:CoherentAuxiliary}
\end{align}
with $q=(t_f/t_0)^{1-r}$. The result in Eq.~\eqref{eq:CorrectionScaling} is a modified version of the identity \eqref{eq:ExcitationDensityScaling2}. The final step is to consider the $t_f\rightarrow\infty$ limit of $\bmcE$. If this limit is finite, then, by the same reasoning that led to Eq.~\eqref{eq:ExcitationDensityScaling3}, we have isolated the leading-order behavior of the nonadiabatic correction $\mathcal{E}(t_f)-\mathcal{E}(0)$ in the $(t_0/t_f)^\zeta$ scaling factor in Eq.~\eqref{eq:CorrectionScaling}. Note that the ramp velocity $v_\mu$ associated with $P(\epsilon,qt_0,\gamma=0,\kappa,q^\beta\dmu_i,T_i=0)$ in the integrand of Eq.~\eqref{eq:CoherentAuxiliary} is fixed at \mbox{$v_\mu=t_0^{-\beta}\dmu_i$}. Moreover, this excitation probability will decrease with increasing $\epsilon$, limiting the range of $\epsilon$ values that contribute significantly to the integral. Based on this, we write
\begin{widetext}
\begin{align}
	\lim_{t_f\rightarrow\infty}\bmcE&=(1-2P^{\rm th}(\Lambda(\dmu_i,0)/T_i))\lim_{q\rightarrow\infty}\int_0^\infty{\rm d}\epsilon\tilde{\rho}(\epsilon)P(\epsilon,qt_0,0,\kappa,q^\beta\dmu_i,0)\nonumber\\
	&=(1-2P^{\rm th}(\Lambda(\dmu_i,0)/T_i))\mathcal{D}(\gamma=0,\kappa,\bar{v}_\mu=t_0^{-\beta}\dmu_i,\bar{v}_T=0).
\end{align}
\end{widetext}
Here the $\mathcal{D}$ excitation density is that of an infinite-time, fixed velocity ramp, as introduced in Sec.~\ref{sec:ConstandSpeedRamps}.
The existence of the limit leading to \mbox{$\mathcal{D}(\gamma=0,\kappa,\bar{v}_\mu=t_0^{-\beta}\dmu_i,\bar{v}_T=0)$} follows from the fact that the unitary dynamics is adiabatic sufficiently far from the critical point where the excitation gap is large. Moving the starting point $\mu_i$ of the ramp further away from $\mu_c$ while keeping the ramp velocity $v_\mu$ fixed therefore eventually leads to the constant nonzero excitation density $\mathcal{D}$. To leading order in $1/t_f$ we then have 
\begin{multline}
	\mathcal{E}(t_f)-\mathcal{E}(0)=(t_0/t_f)^\zeta(1-2P^{\rm th}(\Lambda(\dmu_i,0)/T_i))\quad\\
   \times\mathcal{D}(\gamma=0,\kappa,\bar{v}_\mu=t_0^{-\beta}\dmu_i,\bar{v}_T=0)
\end{multline}
with $\zeta$ as in Eq.~\eqref{eq:CoherentScalingPower}. For a linear ramp with $\beta=1$ this reduces to the well-known expression for the standard Kibble-Zurek protocol \cite{Dziarmaga10}. For $\beta\neq1$ it matches the result for nonlinear ramps in Refs.~\cite{Barankov08,Sen08,DeGrandi10}. 

\bibliographystyle{quantum}
\bibliography{MK}%

\end{document}